\documentclass[a4paper,fleqn,usenatbib]{mnras}
\pdfoutput=1

\usepackage[pdftex]{graphicx,color}
\usepackage[T1]{fontenc}
\usepackage[utf8]{inputenc}
\usepackage{lmodern}

\definecolor{urlblue}{rgb}{0,0,0.9}
\definecolor{linkgreen}{rgb}{0,0.45,0}
\definecolor{linkorange}{rgb}{0.7,0.1,0.0}

\usepackage{amsmath, amssymb}
\usepackage{cuted}
\usepackage[english]{babel}
\usepackage{enumerate}
\usepackage[normalem]{ulem}
\usepackage{enumitem}
\usepackage{multirow}
\usepackage{booktabs}
\usepackage{makecell}

\setlist[enumerate]{wide=0pt, widest=99,leftmargin=\parindent, labelsep=* } 

\definecolor{valecol}{rgb}{0,0.5, 1.}

\definecolor{davidcol}{rgb}{0,0, 0.75}

\graphicspath{{./figs/}}

\AtBeginDocument{\hypersetup{
linkcolor=linkgreen,
citecolor=linkorange,
urlcolor=urlblue}}

\def\L{\mathcal{L}}
\def\E{\mathcal{E}}

\def\d{{\rm d}}

\title[New (inverse) distance ladder]{A new method to build the  (inverse) distance ladder}

\author[Camarena and Marra]{
David Camarena$^{1,2}$
and Valerio Marra$^{3}$
\\
$^{1}$PPGCosmo, Universidade Federal do Espírito Santo, 29075-910, Vitória, ES, Brazil\\
$^{2}$Institut für Theoretische Physik, Universität Heidelberg, Philosophenweg 16, 69120 Heidelberg, Germany\\
$^{3}$Núcleo de Astrofísica e Cosmologia \& Departamento de Física, Universidade Federal do Espírito Santo, 29075-910, Vitória, ES, Brazil
}

\date{Accepted XXX. Received YYY; in original form ZZZ}

\pubyear{20XX}

\begin{document}
\label{firstpage}
\pagerange{\pageref{firstpage}--\pageref{lastpage}}

\maketitle

\begin{abstract}

The cosmic distance ladder is the succession of techniques by which it is possible to determine distances to astronomical objects.
Here, we present a new method to build the cosmic distance ladder, going from local astrophysical measurements to the CMB.
Instead of relying on high-redshift cosmography in order to model the luminosity-distance relation and calibrate supernovae with BAO, we exploit directly the distance-duality relation $d_L = (1+z)^2 d_A$---valid if photon number is conserved and gravity is described by a metric theory.
The advantage is that the results will not depend on the parametrization of the luminosity-distance relation at $z>0.15$: no model is adopted in order to calibrate BAO with supernovae.
This method yields local measurements of the Hubble constant and deceleration parameter. Furthermore, it can directly assess the impact of BAO observations on the strong 4--5$\sigma$ tension between local and global $H_0$.
Using the latest supernova, BAO and CMB observations, we found a consistently low value of $q_0$ and strong inconsistency between angular-only BAO constraints and anisotropic BAO measurements, which are, or not, in agreement with CMB depending on the kind of analysis (see Table~\ref{tab:full_pos}).
We conclude that, in order to understand the reasons behind the $H_0$ crisis, a first step should be clarifying the tension between angular and perpendicular anisotropic BAO  as this will help  understanding if new physics is required at the pre-recombination epoch or/and during the dark energy era.

\end{abstract}


\section{Introduction}\label{sec:intro}

The cosmic distance ladder is the succession of techniques by which it is possible to determine distances to astronomical objects.
No single method can measure distances at all ranges encountered in astronomy and each technique, or rung of the ladder, provides information that can be used to determine the distances at the next higher rung.
The cosmic distance ladder provides a model-independent way to constraint the Hubble constant at local scales. The latest and most accurate measurement is $H_0 = 73.5 \pm 1.4$ km s$^{-1}$ Mpc$^{-1}$ \citep[][SH0ES]{Reid:2019tiq}  which is at $4.2\sigma$ tension with the cosmological constraint $H_0 = 67.36 \pm 0.54$ km s$^{-1}$ Mpc$^{-1}$ from CMB  \citep[][Planck Collaboration]{Aghanim:2018eyx}.
The tension reaches the $4.5\sigma$ if one adopts the determination $H_0 = 75.35 \pm 1.68$ km s$^{-1}$ Mpc$^{-1}$
by \citet{Camarena:2019moy} which uses only local observations and assumes only the cosmological principle, that is, large-scale homogeneity and isotropy.
Because the CMB analysis is based on the flat $\Lambda$CDM model, this tension could point towards the existence of new physics beyond the standard model.
Indeed, measurements using other methods or calibrations confirm this tension and it is difficult that a single systematic effect could explain it away.
See \citet{Verde:2019ivm}  for  a detailed review of the present status and observational effort to determine $H_0$.

Here, first we will review how the cosmic distance ladder can be extended from our galaxy to the CMB.
This allows one to use it as an ``inverse distance ladder'', that is, to calibrate the supernovae via CMB and BAO instead of Cepheids and geometrical distances \citep{Cuesta:2014asa,Aubourg:2014yra,Verde:2016ccp,Lemos:2018smw,Feeney:2018mkj,Macaulay:2018fxi,Tutusaus:2018ulu,Taubenberger:2019qna}.
Although this method allows one to avoid possible unknown systematics, most of the approaches requires a fiducial model to convert redshifts into distances \citep{Cuesta:2014asa,Aubourg:2014yra,Verde:2016ccp,Lemos:2018smw,Taubenberger:2019qna}. A few cases have been built using a cosmographic expansion \citep{Feeney:2018mkj,Macaulay:2018fxi}, 
which could be problematic at $0 \lesssim z \lesssim 2$ \citep{Zhang:2016urt}.
Most importantly, the standard approach cannot provide a truly local value of $H_0$ as supernovae beyond $z=0.15$ are used. In other words, the cosmographic method forces correlation between $H_0$ and the shape of the luminosity-distance relation at $0\lesssim z \lesssim 2$.

In order to overcome this issue, we propose a new way to build the cosmic distance ladder. Instead of using a cosmographic model to calibrate supernovae and BAO at the same time, we will bin supernovae at $z  > 0.15$ such that the final luminosity distance can be directly compared with the BAO distance via the distance-duality relation $d_L = (1+z)^2 d_A$---valid if photon number is conserved and gravity is described by a metric theory.
In other words, our binning procedure is used to calibrate supernovae and BAO observations and not to reconstruct the luminosity distance \citep[see, e.g.,][and references therein]{Sapone:2014nna,Cao:2017gfv,Lemos:2018smw,Lyu:2020lwm}. 
Indeed, the crucial point is that no assumption is made regarding the functional form of the luminosity distances or the dynamical behavior of the universe beyond~$z = 0.15$.
The potential bias of the cosmographic method is absent in the calibration that we propose here. Also, this method can identify in a direct way the role that BAO measurements play on the Hubble tension.

This paper is organized as follows.
In Section~\ref{ladder} we describe the cosmic distance ladder, going from local astrophysical probes to the CMB. In Section~\ref{sec:bin} we present our new method to calibrate supernovae with BAO and in Section~\ref{analyses} we will build three different distance ladders: one that uses only supernova and BAO data, one the uses supernova, BAO and CMB data (the inverse distance ladder) and one that uses supernova, BAO and local astrophysical data (the extended distance ladder).
Then, we will present our results in Section~\ref{sec:resul}, which we compare with previous distance ladders (and similar approaches) in Section~\ref{sec:discu}. The conclusions of Section~\ref{sec:conclu} are followed by the three Appendixes \ref{ap:cosmography}-\ref{ap:disjoint}.

\section{The cosmic distance ladder}\label{ladder}

\begin{figure*}
\centering
\includegraphics[width=.98 \textwidth]{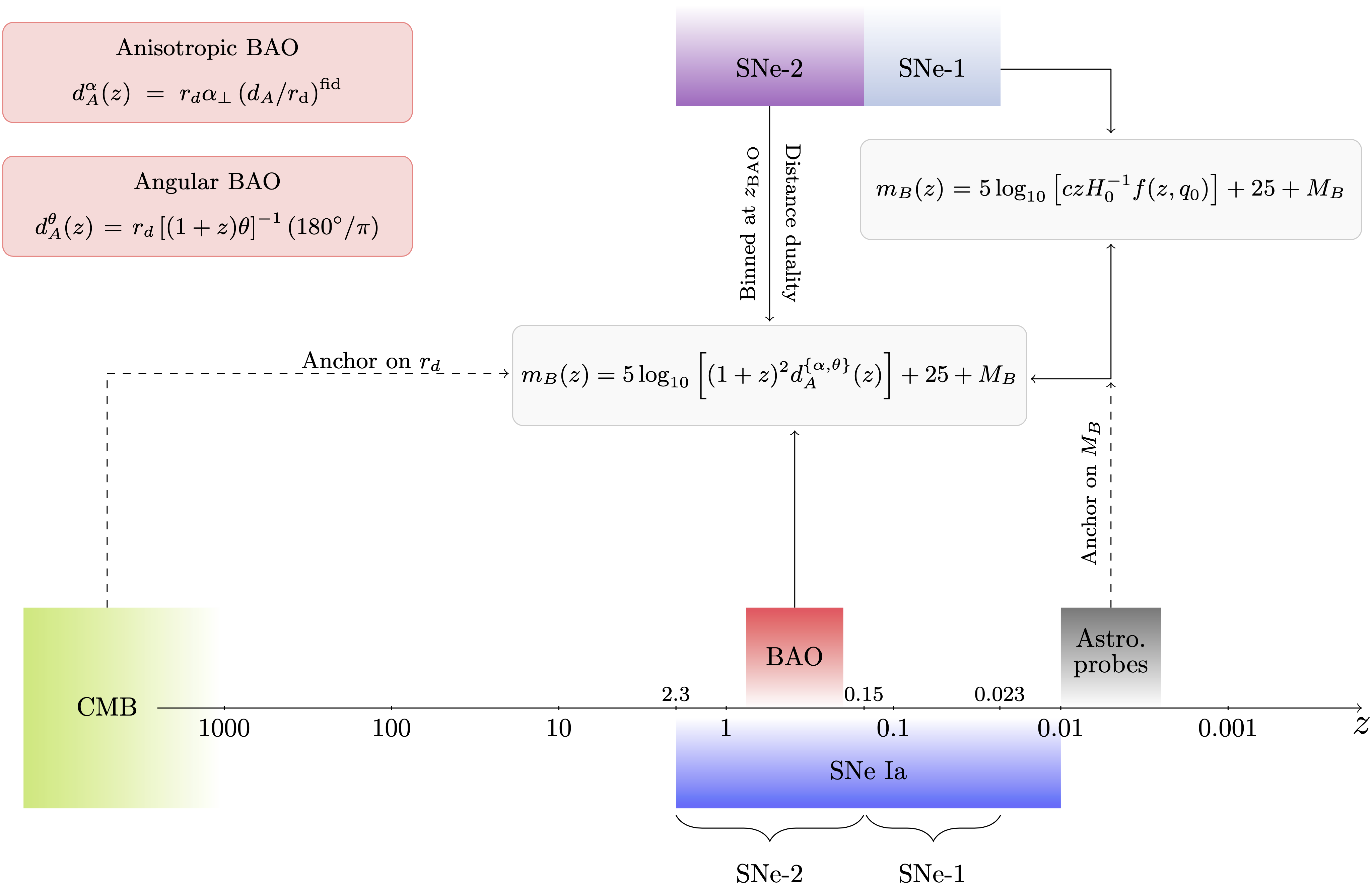}
\caption{Flowchart of our new method to build the cosmic distance ladder.
The analysis starts with choosing a prior or anchor (dashed line), which is then propagated via supernovae and BAO distances to its opposite side.
This leads to constraints on $H_0$ or $r_d$. It is also possible to use  BAO and supernovae  to constrain $r_d h$ without any prior information. Supernovae are calibrated with BAO via the distance-duality relation in a model-independent way.}
\label{fig:diag}
\end{figure*}

The cosmic distance ladder is the succession of techniques by which it is possible to determine distances to astronomical objects.
Indeed, no single method can measure distances at all ranges encountered in astronomy but rather is limited to a specific range.
Each technique, or rung of the ladder, provides information that can be used to determine the distances at the next higher rung.
In this Section, we will review how the cosmic distance ladder is extended from our galaxy to the CMB.
The diagram of Figure~\ref{fig:diag} summarizes the method developed in this work.

\subsection{Local astrophysical probes ($z\lesssim 0.01$)} \label{astrosec}

The first rung of the ladder is meant to calibrate supernovae~Ia (SNe) in nearby galaxies starting from direct distance measurements via parallax.
The chain of intermediate calibrations is complex, involving astrophysical correlations and observables. See \citet{Riess:2016jrr} for details.

In \citet{Camarena:2019moy} we presented a method to compress the first rungs of the ladder (see Figure~\ref{fig:diag}) into an effective calibration prior on the absolute magnitude $M_B$ of supernovae Ia.
The calibration prior on $M_B$ relative to \citet{Reid:2019tiq} is:
\begin{align} \label{priorM}
M_B = -19.2334 \pm 0.0404 \text{ mag} \,.
\end{align}
It is important to stress that this ``astro-prior'' is cosmology independent.
Also, the use of the astro-prior avoids the double counting of low-redshift supernovae that happens in analyses that use the local $H_0$ determination by \citet{Riess:2019cxk}.
Indeed, in the standard analysis, low-$z$ supernovae are used twice: once for the $H_0$ determination and once when constraining the cosmological parameters, thus this double counting could produce a bias. The advantage of adopting  the calibration prior is that supernovae are used only once.
See also \citet{Benevento:2020fev} where the astro-prior is used in the context of models that feature late-time dark energy transitions.

\subsection{Supernovae Ia ($0.01<z< 2.3$)} \label{sec:SNe}

Type Ia supernovae are standardizable candles.
After the standardization procedure, the supernova absolute magnitude $M_B$, although unknown, is expected to remain constant in redshift \citep{Scolnic:2017caz}, and this allows us to use supernovae to connect observables at low ($z\sim 0.01$) and high ($z\sim 2$) redshifts.

In particular, as far as the cosmic distance ladder is concerned, supernovae can be used in two complementary ways:
\begin{enumerate}

\item to transport the astro-prior to the redshift of BAO observations (see below) and so obtain a calibration of the sound horizon at drag epoch $r_d$,

\item to calibrate $M_B$ with BAO via a CMB prior on $r_d$ and transport it to low redshift.

\end{enumerate}

For a supernova at redshift  $z$, the apparent magnitude $m_B$ is given by:
\begin{equation}
m_B = 5\log_{10}\left[\frac{d_L(z)}{1 \text{Mpc}} \right] + 25 + M_B \,, \label{eq:mB}
\end{equation}
where $d_L(z)$ is the luminosity distance and $M_B$ the absolute magnitude.
Here, we use the Pantheon catalog, which features 1048 Supernovae Ia in the redshift range $0.01 \le z \le 2.3$ \citep{Scolnic:2017caz}.

We will not consider supernovae at $z<0.023$ as at these small scales the local structure significantly perturbs the FLRW metric \citep{Camarena:2018nbr}.
We will split the remaining supernovae into two groups, see Figure~\ref{fig:diag}:
\begin{enumerate}[label=(1)]

\item \textbf{SNe-1}: supernovae in the range $0.023 \le  z \le 0.15$, which will be used to determine local $H_0$, see below.
The superscript $^*$ denotes SNe-1 data: observed redshift $z^*_i$, observed apparent magnitude $m^*_{B,i}$ and covariance matrix $\Sigma^{*}$ (including systematics).

\item \textbf{SNe-2}: supernovae with $z > 0.15$, which will be used to connect with BAO and extend the ladder.
SNe-2 data is denoted by quantities without the superscript~$^*$.

\end{enumerate}
Although we split SNe in two groups,  our analysis considers correlations between SNe-1 and SNe-2. The opposite case and its implications are discussed in Appendix~\ref{ap:disjoint}.

\subsubsection{Determination of local $H_0$} \label{locdet}

Any calibration of $M_B$, from CMB or astrophysical probes, directly yields a measurement of the local Hubble constant. 
The determination of $H_0$ is achieved by fitting a model-independent cosmographic expansion to local supernovae (SNe-1), that is, supernovae in the range $0.023 \le  z \le 0.15$ \citep{Riess:2016jrr}, where the minimum redshift is large enough in order to reduce cosmic variance \citep[][an references therein]{Camarena:2018nbr,Marra:2013rba,Macpherson:2018akp} and the maximum redshift is small enough in order to reduce the dependence on cosmology and classify the measurement as ``local.''
The theoretical apparent magnitude is given via the cosmographic approximation:
\begin{align}
&m^{\rm cg}_B  =  5 \log_{10} \left [\frac{c z H_0^{-1}}{\text{1 Mpc}} \, f  (z,q_0 ) \right] + 25 + M_B \,, \label{eq:mB_cosmography} \\
&f(z,q_0) = 1 + \frac{1-q_0}{2} z + O(z^2) \,, \nonumber
\end{align}
where $q_0$ is the current value of the deceleration parameter and second-order terms have been neglected.\footnote{
As for SNe-1 it is $z<0.15$, the weighted error from neglecting the second-order correction is only 0.2\% \cite[see][figure 1]{Camarena:2019moy}. For a more detailed discussion, see Appendix~\ref{ap:cosmography}.} 
The Hubble constant is then obtained via Bayesian inference: 
\begin{align}
&f(H_0,q_0, M_B | \text{SNe-1}) = \frac{f(H_0)f(q_0) f(M_B) \L }{\E} \,, \label{h0loc} \\
&f(H_0 | \text{SNe-1}) = \int \d M_B \d q_0  f(H_0, q_0,M_B  | \text{SNe-1}) \,,
\end{align}
where the likelihood is given by:
\begin{align}
&\L(\text{SNe-1} | H_0, q_0,M_B) = |2 \pi \Sigma^*|^{-1/2}  e^{-\frac{1}{2}\chi^{2}_\text{SNe-1}(H_0, q_0,M_B)} \,, \\
&\chi^{2}_\text{SNe-1}= \{ m^*_{B,i}  - m_B^{\rm cg}(z^*_i) \}  \Sigma^{*-1}_{ij}  \{ m^*_{B,j}  - m_B^{\rm cg}(z^*_j) \} \,.
\end{align}
In equation~\eqref{h0loc} $f(M_B)$ is the calibration of $M_B$ and $f(H_0)$ is a flat uninformative prior. Regarding $f(q_0)$, three choices have been adopted so far:
\begin{align}
&f(q_0) = \delta (q_0 + 0.55)  \hspace{0.6cm} \text{\citet{Reid:2019tiq}}, \label{deltaq} \\
& f(q_0) = \mathcal{N}(-0.55,1)  \hspace{0.57cm} \text{\citet{Feeney:2017sgx}}, \\
&f(q_0) = \text{constant}  \hspace{1.07cm} \text{\citet{Camarena:2019moy}} \,, \label{uniq}
\end{align}
where $\delta$ is the Dirac delta function and $\mathcal{N}$ a normal distribution.
The approach by \citet{Camarena:2019moy} has the advantage that it delivers a determination of $H_0$ that only assumes the FLRW metric as it does not fix $q_0$ to the standard-model value of $-0.55$.

Adopting the prior of eq.~\eqref{priorM}, one obtains the values given in \citet{Reid:2019tiq} and \citet{Camarena:2019moy}, using the priors of eq.~\eqref{deltaq} and \eqref{uniq}, respectively:
\begin{align}
&H^{\rm Re19}_0 = 73.5 \pm 1.4 \text{ km s}^{-1} {\rm Mpc}^{-1} \,\text{,} \label{reid19} \\
&H_0^{\rm CM} = 75.35 \pm 1.68 \text{ km s}^{-1} {\rm Mpc}^{-1} \,\text{.} \label{cm19}
\end{align}
%

\subsection{Baryon acoustic oscillations} \label{sec:BAO}

The detection of baryon acoustic oscillations yields the primordial sound horizon $r_d$ at the redshift of the galaxy sample used for the analysis.
This makes BAO a standard ruler.
BAO measurements have a potential higher accuracy as compared to supernovae.
First, they involve almost-linear well-understood physics as opposed to the complicated astrophysical processes that lead to supernovae' explosions.
Second, the determination of $r_d$ is calibrated, although degenerate with cosmological quantities such as the angular distance $d_A$ and the Hubble rate $H(z)$, see equations (\ref{eq:alphas}, \ref{eq:thetabao}). The supernovae' magnitude $M_B$  has instead to be calibrated with external data.
Finally, future galaxy catalogs will increase dramatically the quality of BAO measurements; see, for instance, \citet{Amendola:2016saw,Aghamousa:2016zmz,Benitez:2014ibt} and~\citet{Marra:2019lyc} for a recent review.

Usually, BAO measurements are used together with CMB data. Indeed, the latter tightly constrain $r_d$ so that the degeneracies between $r_d$ and $d_A, H$ are broken. BAO determinations can then constraints the cosmological parameters that enter $d_A$ and $H$.

Here, instead, we use BAO measurements to connect supernovae to CMB observations. As we discuss below, we calibrate the angular distance from BAO with the luminosity distance from supernovae in a cosmology-independent way.
In this way one can propagate the CMB constraint on $r_d$ into a constraint on $M_B$ (inverse ladder) or propagate the astro-prior on $M_B$ into a constraint on $r_d$ (extended ladder), see Figure~\ref{fig:diag}.

In order to obtain the angular diameter distance from BAO we will use the results of anisotropic \citep{Alam:2016hwk} and angular \citep{Carvalho:2015ica,Alcaniz:2016ryy,Carvalho:2017tuu} BAO analyses.
These measurements will not be used together as they were obtained from the same BOSS galaxy catalog.

\subsubsection{Anisotropic BAO} \label{subsec:BAO_ani}

Anisotropic BAO analyses determine the BAO feature  along both the line-of-sight and transverse directions.
Here, we use the consensus results from the final BOSS-DR12 sample \citep{Alam:2016hwk}, which comprises 1.2 million massive galaxies over an effective volume of $18.7$ Gpc$^3$.  
The analysis by \citet{Alam:2016hwk} adopts a fiducial model (a flat $\Lambda$CDM model with $\Omega_m = 0.31$, $h=0.676$, $\Omega_b h^2 = 0.022$, $\sigma_8 = 0.8$ and $n_s =0.97$) to convert redshifts into distances.
However, the final result is not biased because
deviations from this fiducial model are allowed in the final fit.%
\footnote{This is valid if the adopted model is ``close'' enough to the trial model. See \citet{Anselmi:2015dha,Marra:2018zzo} for alternative analyses.}
To this end, instead of fitting the BAO peak position,
the analysis provides a measurement of the dilation parameters $\alpha_{\perp}$ and $\alpha_\parallel$, which are defined according to:
\begin{gather}
\alpha_{\perp} \equiv \frac{d_A(z)}{r_d }\frac{r_d^{\text{fid}}}{d_A(z)^{\text{fid}}}
\phantom{belezapura}
\alpha_{\parallel} \equiv \frac{H(z)}{H(z)^{\text{fid}}}\frac{r_d}{r_d^{\text{fid}}} \,. \label{eq:alphas}
\end{gather} 
The dilation parameters quantify how much the BAO peak can shift from the peak in the fiducial model \cite[see][and references therein]{Alam:2016hwk}.

\citet{Alam:2016hwk} reports distance measurements at three effective redshifts: $0.38 $, $0.51$ and $0.61$. However, in order to avoid overlapping redshift bins and simplify the supernova binning discussed in section~\ref{sec:bin}, we only consider the determinations at $0.38 $ and $0.61$.
Finally, as we need the angular distance, we only consider the perpendicular dilation parameter $\alpha_\perp$. Hereafter, for the sake of simplicity we refer to the perpendicular anisotropic BAO as anisotropic BAO.
The data is shown in Table~\ref{tab:BAO_ani}. 

\begin{table}
\centering
\begin{tabular}{llllc}
\hline\hline
Catalog & $z_{\text{bao}}$ & $ \alpha_\perp(z)$  & $\sigma_{\alpha}$ & Reference \\ \hline
BOSS-DR12 & $0.38$ & $0.993 $   & $0.0146$ & \cite{Alam:2016hwk} \\ 
 & 0.61 & $0.990$  & $0.0138$ &  \\ 
\hline\hline
\end{tabular}
\caption{Anisotropic BAO data. The correlation between the two determinations is 0.2.}
\label{tab:BAO_ani}
\end{table}

\subsubsection{Angular BAO} \label{subsec:BAO_ang}

It is possible to measure the angular BAO scale in a model-independent way if thin-enough redshift bins are used~\citep{Sanchez:2010zg}.
In this case the measurements constrain the combination:
\begin{equation}
	\theta(z)=\frac{r_d}{d_A(z) (1+z)} \left(\frac{180^\circ}{\pi}\right) \,. \label{eq:thetabao}
\end{equation}
The data is presented in table~\ref{tab:BAO_ang}.

\begin{table}
\centering
\begin{tabular}{llllc}
\hline\hline
Catalog & $z_{\rm{bao}}$    & $\theta(z)$ & $\sigma_{\theta}$ & Reference \\
\hline
BOSS-DR7          & 0.235        & 9.06        & 0.23        & \cite{Alcaniz:2016ryy} \\
BOSS-DR7          & 0.365        & 6.33        & 0.22        & \cite{Alcaniz:2016ryy} \\
BOSS-DR10         & 0.450        & 4.77        & 0.17        & \cite{Carvalho:2015ica} \\
BOSS-DR10         & 0.470        & 5.02        & 0.25        & \cite{Carvalho:2015ica} \\
BOSS-DR10         & 0.490        & 4.99        & 0.21        & \cite{Carvalho:2015ica} \\
BOSS-DR10         & 0.510        & 4.81        & 0.17        & \cite{Carvalho:2015ica} \\
BOSS-DR10         & 0.530        & 4.29        & 0.30        & \cite{Carvalho:2015ica} \\
BOSS-DR10         & 0.550        & 4.25        & 0.25        & \cite{Carvalho:2015ica} \\
BOSS-DR11         & 0.570        & 4.59        & 0.36        & \cite{Carvalho:2017tuu} \\
BOSS-DR11         & 0.590        & 4.39        & 0.33        & \cite{Carvalho:2017tuu} \\
BOSS-DR11         & 0.610        & 3.85        & 0.31        & \cite{Carvalho:2017tuu} \\
BOSS-DR11         & 0.630        & 3.90        & 0.43        & \cite{Carvalho:2017tuu} \\
BOSS-DR11         & 0.650        & 3.55        & 0.16        & \cite{Carvalho:2017tuu} \\
\hline\hline
\end{tabular}
\caption{Angular BAO data.}
\label{tab:BAO_ang}
\end{table}

\subsection{Cosmic microwave background}

The CMB tightly constrains the sound horizon at the drag epoch and is used to anchor the inverse distance ladder.
Here, we adopt the constraint from the TT, TE, EE + lowE + lensing Planck 2018 analysis \citep{Aghanim:2018eyx}:
\begin{equation} \label{cmbprior}
r_d = 147.09 \pm 0.26 \text{ Mpc} \,.
\end{equation}
%

\subsection{Other probes}

Additional observations may be used to improve the accuracy of the cosmic distance ladder.
For example, one may use luminosity distance determinations from standard sirens in order to calibrate supernovae and/or BAO \citep{Gupta:2019okl}.
The effect of these extra probes is to reduce statistical errors and biases due to systematics.

\section{New method to calibrate Type Ia supernovae with BAO} \label{sec:bin}

Here, we propose a new method to calibrate supernovae with BAO.
The idea is to exploit directly the distance-duality relation $d_L = (1+z)^2 d_A$ --  valid if photon number is conserved and gravity is described by a metric theory -- without having to use a cosmological or cosmographic model.
To achieve the latter we will bin the supernovae SNe-2 at the effective redshift $z_{\text{bao}}$ of the BAO observations, so that luminosity and angular distances can be directly compared.

In the standard inverse ladder approach a cosmographic model is used to obtain the luminosity distance, which is fitted to  the full supernova dataset spanning $0\lesssim z \lesssim 2$.
In other words, the value of $d_L(z_{\text{bao}})$ depends on the global fit of a parametrized luminosity distance to all the supernovae, and this remains true even if one marginalizes over the model parameters. 
This potential bias is absent in the calibration that we present here.

We emphasize that the novelty of our method lies in avoiding a cosmography expansion beyond local scales, providing the opportunity to constraint local parameters using early universe priors (and vice-versa).
Since the distance-duality is explicit used at the same redshift for binned supernovae and BAO distances, our results do not depend of the functional form of the luminosity distance nor of the dynamical behavior of the universe beyond $z = 0.15$.

The method presented in the next sections works  for a supernova catalog like the Pantheon~\citep{Scolnic:2017caz} whose covariance matrix does not dependent on nuisance parameters, and it should be generalized if one wishes to use a catalog like JLA~\citep{Betoule:2014frx}.

\subsection{Weighted average}\label{subsec:weighted}

A supernova  catalog like Pantheon 
provides observed redshifts and apparent magnitudes $\left\lbrace z_i, m_{Bi} \right\rbrace$ together with the covariance matrix $C_{ij}$ (including systematic and statistical errors), where $i,j=\left\lbrace 1,\dots,N\right\rbrace$. 
In order to produce a binned catalog with $n$ bins we have to transform the $N$
data points of the full catalog 
into the new points $\left\lbrace {z}_a,{m_{Ba}} \right\rbrace$, where  $a=\left\lbrace 1,\dots,n\right\rbrace$. The binning process has to be such that $\left\lbrace {z}_a,{m_{Ba}} \right\rbrace$ are a good representation of $\left\lbrace z_i, m_{Bi} \right\rbrace$ in the $a$-th bin. In a statistical language, the point ${x}_a$ (here $x$ represents $z$ or $m_B$) minimizes the function \citep{Schmelling:1994pz}:
\begin{equation}
\chi^2_a= \sum_{ij}^{n_a} \left\lbrace x_i^{\rm a} - {x}_a \right\rbrace (C_{\rm a}^{-1})_{ij} \left\lbrace x_j^{\rm a} - {x}_a \right\rbrace \,, \label{eq:chi2bin}
\end{equation}
where $n_a$ is the number of supernovae in the $a$-th bin and the super/underscript $^{\rm a}$ 
indicates that the quantity belongs to the $a$-th bin $[z^{\rm a}_l,z^{\rm a}_r] $.
Thus, it can be demonstrated that:
\begin{gather}
{z}_a = \frac{\sum\limits_{i,j=1}^{n_a}(C_{\rm a}^{-1})_{ij} z^{\rm a}_i}{\sum\limits_{i,j=1}^{n_a}(C_{\rm a}^{-1})_{ij}} 
\phantom{ciaociao}
{m_{Ba}} = \frac{\sum\limits_{i,j=1}^{n_a}(C_{\rm a}^{-1})_{ij} m^{\rm a}_{Bi}}{\sum\limits_{i,j=1}^{n_a}(C_{\rm a}^{-1})_{ij}} \,. \label{eq:zmB_bin} 
\end{gather}
While we will bin only SNe-2, formally the results above are valid also for SNe-1: for supernovae that are not binned it is $n_a=1$, $z_a = z^{\rm a}_1$ and $m_{Ba}=m^{\rm a}_{B1}$.
In this way one has $N= \sum_a n_a$.

\subsection{Binned covariance matrix}

In order to conclude the supernova binning we need to compute the corresponding binned covariance matrix $C_{ab}$ with $a,b=\left\lbrace 1,\dots,n\right\rbrace$.
In order not to neglect correlations it is important to consider all the catalog ($N= \sum_a n_a$), that is, to bin the full covariance matrix $C_{ij}$.

Now, it is clear that the quantities $\left\lbrace {z}_a,{m_{Ba}} \right\rbrace$ are a linear transformation of $\left\lbrace z_i,m_{Bi} \right\rbrace$. Therefore, to find the binned covariance matrix $C_{ab}$, we can use first-order uncertainty propagation, which is exact in this linear case.
We define then the Jacobian:
\begin{equation}
J_{ai}=\frac{\partial {z}_a}{\partial z_i} =\left\lbrace \!\!\!\!\!
\begin{array}{cl}
\frac{\sum\limits_{j=1}^{n_a}(C_{\rm a}^{-1})_{ji}}{\sum\limits_{j,k=1}^{n_a}(C_{\rm a}^{-1})_{jk}} & \text{if } z_{i} \in [z^{\rm a}_l,z^{\rm a}_r] \\
0 & \text{if }  z_{i} \notin [z^{\rm a}_l,z^{\rm a}_r] 
\end{array}
\right. \,,
\label{eq:jaco}
\end{equation} 
where equation \eqref{eq:zmB_bin} was used.
Note that for supernovae that are not binned (the ones belonging to SNe-1), it is $J_{ai}=\delta_{ai}$.
The binned covariance matrix is then:
\begin{equation}
C_{ab}= J_{ai} \, C_{ij} \, J_{bj}  \,. \label{eq:covbin}
\end{equation} 
It is important to mention that our binning process uses the full covariance matrix, which includes systematic and statistic errors.

Note that in the equations above the bins are not overlapping so that the averaging process in a given bin does not affect the other bins, see equation~\eqref{eq:zmB_bin}.
This is the motivation for using only the two non-overlapping BOSS measurements of Table~\ref{tab:BAO_ani} so that the supernova bins are also non-overlapping.

\subsection{Binning applied to Pantheon}

\begin{table*}
\begin{center}
\renewcommand{\arraystretch}{1.3}
\begin{tabular}{|c|c|c|c|c|c|c|c|}
\hline
\hline
$a$-th bin & BAO range & $z_{\rm{bao}}$ & SNe-2 bin range & ${z}_a$ & $n_a$ & $m_{Ba}$ & $\overline m_{Ba}(z_{\rm{bao}})$ \\ \hline
$1$ & $[0.20,0.50]$ & $0.38$ & $[0.3088,0.49737]$ &$0.380059$ & 183 & 22.19407 & 22.19366 \\ \hline
$2$ & $[0.50,0.75]$ & $0.61$ &  $[0.5109,0.74909]$  & $0.610372$ & 108 & 23.42215 & 23.42057  \\
\hline
\hline
$a$-th bin & $[z^{\rm{bao}}_l,z^{\rm{bao}}_r]$ & $z_{\rm{bao}}$ & $[z^{\rm a}_l,z^{\rm a}_r]$ & ${z}_a$ & $n_a$ & $m_{Ba}$ & $\overline m_{Ba}(z_{\rm{bao}})$  \\ \hline
$1$ & $[0.20,0.27]$ & $0.235$ & $[0.1998,0.26828]$ &$0.23496$ & 162 & 20.99119 & 20.9915 \\ \hline
$2$ & $[0.34,0.39]$ & $0.365$ & $[0.34701,0.38887]$  & $0.36504$ & 56 & 22.10539 & 22.10517 \\ \hline
$3$ & $[0.44,0.46]$ & $0.45$ &  $[0.44255,0.4604]$  & $0.44965$ & 10 & 22.60076 & 22.60360 \\ \hline
$4$ & $[0.465,0.475]$ & $0.47$ &  $[0.46664,0.47572]$  & $0.47064$ & 7 & 22.77238 & 22.76838 \\ \hline
$5$ & $[0.48,0.50]$ & $0.49$ &  $[0.48072,0.50074]$  & $0.49046$ & 6 & 22.89645 & 22.89485 \\ \hline
$6$ & $[0.505,0.515]$ & $0.51$ &  $[0.50309,0.51434]$  & $0.5097$ & 9 & 22.96283 & 22.96356 \\ \hline
$7$ & $[0.525,0.535]$ & $0.53$ &  $[0.52278,0.53378]$  & $0.52962$ & 4 & 23.01131 & 23.01307 \\ \hline
$8$ & $[0.545,0.555]$ & $0.55$ &  $[0.54539,0.55260]$  & $0.54952$ & 5 & 23.10485 & 23.11065 \\ \hline
$9$ & $[0.565,0.575]$ & $0.57$ &  $[0.56475,0.56475]$  & $0.56475$ & 1 & 23.2891 & 23.2884 \\ \hline
$10$ & $[0.585,0.595]$ & $0.59$ &  $[0.58873,0.59185]$  & $0.59032$ & 3 & 23.28571 & 23.28203 \\ \hline
$11$ & $[0.605,0.615]$ & $0.61$ &  $[0.60825,0.6157]$  & $0.61098$ & 5 & 23.52225 & 23.52964 \\ \hline
$12$ & $[0.625,0.635]$ & $0.63$ &  $[0.62723,0.63771]$  & $0.63075$ & 6 & 23.37277 & 23.36319  \\ \hline
$13$ & $[0.64,0.66]$ & $0.65$ &  $[0.64364,0.66206]$  & $0.65046$ & 4 & 23.62446 & 23.61859 \\ \hline
\hline
\end{tabular}
\caption{Result of the binning process for the anisotropic BAO data of Table~\ref{tab:BAO_ani} (top) and for the angular BAO data of Table~\ref{tab:BAO_ang} (bottom).}
\label{tab:BAO_bin}
\end{center}
\end{table*}

Now, we apply the method of the previous sections to the Pantheon dataset.
As said earlier, we wish to obtain the luminosity distance of supernovae at $z_{a}=z_{\rm{bao}}$, using SNe-2.
In order to achieve the latter we have to find the appropriate bin $[z^{\rm a}_l,z^{\rm a}_r]$.
The redshift ranges $[z^{\rm{bao}}_l,z^{\rm{bao}}_r]$ that were used in the BAO analyses are given in Table~\ref{tab:BAO_bin}.
We adopt the following algorithm:
\begin{enumerate}[label=\arabic*.,leftmargin=.5\parindent]
	\item We choose $z^{\rm a}_l \simeq z^{\rm{bao}}_l$, where $z^{\rm a}_l$ is the redshift of the supernova that is closest to $z^{\rm{bao}}_l$;
	\item We choose the supernova whose redshift $z^{\rm a}_r \in [z^{\rm{bao}}_l,z^{\rm{bao}}_r]$ minimizes the cost function $f=|z_{a}-z_{\rm{bao}}|$. This will give $ [z^{\rm a,1}_l,z^{\rm a,1}_r] $ together with the cost $f_{1}$.
	\item We repeat step 1 by choosing $z^{\rm a}_r \simeq z^{\rm{bao}}_r$ and step 2 by choosing the $z^{\rm a}_l$ that minimizes $|z_{a}-z_{\rm{bao}}|$.  This will give $ [z^{\rm a,2}_l,z^{\rm a,2}_r] $ together with the cost $f_{2}$.
	\item Choose the bin $[z^{\rm a}_l,z^{\rm a}_r]= [z^{\rm a,i}_l,z^{\rm a,i}_r] $ with the lowest cost~$f_{i}$.
\end{enumerate}
Note that this strategy ensures that supernova and BAO observations probe the same redshift interval, that is, the same physics. Note that no model was used.

Table~\ref{tab:BAO_bin} shows the result of the algorithm above.
One can note that, as expected, $z_a$ is not exactly equal to $z_{\rm{bao}}$ and, therefore, $m_{Ba}$ slightly differs from the desired  $\overline m_{Ba}(z_{\rm{bao}})$.
In order to correct for this small potential bias, we exploit the phenomenological fact that the distance modulus is well approximated by a piece-wise linear function of log$(z)$, see \citet[][Eq.~(E.1) and Fig.~E.1]{Betoule:2014frx} and \citet[][Fig.~11]{Scolnic:2017caz}.
We then correct $m_{Ba}$ according to:
\begin{equation}
	\overline m_{Ba}(z_{\rm{bao}})= m_{Ba} + \frac{\Delta m_B}{\Delta \log z} \left( \log z_{\rm{bao}} - \log {z}_a \right) \,, \label{eq:mB_logz}
\end{equation}
where $\Delta m_B = m_B(z_a)-m_B(z_{a+1})$ and $\Delta \log z =\log z_a - \log z_{a+1} $.
This relation should be particularly accurate owing to the fact that $|{z}_a - z_{\rm{bao}}| \ll 1$.
The (negligible) correction is given in the last column of Table~\ref{tab:BAO_bin}.
In the following, we will omit the over-bar in $\overline m_{Ba}$. Similarly, when appropriate, $z_a$ will mean $z_{\rm{bao}}$.

In order to validate our binning method, we have produced, starting from the full Pantheon dataset, a binned catalog similar to one provided by \citet{Scolnic:2017caz}.
In Appendix~\ref{ap:vsPantheon}
we perform a Bayesian analysis of a $w$CDM model using the two compressed datasets.
The result shows that the two binned catalogs are indistinguishable and provide the same constraints over the cosmological parameters. 

Finally, we should mention that the method of binning cosmological observables has already been used in cosmology with the aim of reconstructing, for instance, cosmological distances in a model independent way \citep[see, e.g.,][and references therein]{Sapone:2014nna,Cao:2017gfv}.

\subsection{Calibrating supernovae with BAO}

We will now discuss how the calibration of supernovae with BAO (and vice-versa) is carried out.
Using the distance-duality relation $d_L = (1+z)^2 d_A$,
the anisotropic and angular BAO determinations of equations (\ref{eq:alphas}--\ref{eq:thetabao}) can be cast in terms of the luminosity distance according to:
\begin{align}
d_L^{\alpha}(z_{\rm{bao}}) &= \frac{\alpha_{\perp}^i r_d }{r_d^{\text{fid}}} (1+z_{\rm{bao}})^2 d_A^{\text{fid}}(z_{\rm{bao}}) \,, \label{eq:dL_ani} \\
d_L^{\theta}(z_{\rm{bao}}) &= \frac{(1+z_{\rm{bao}}) r_d }{\theta_i} \left( \frac{180^\circ}{\pi} \right) \,. \label{eq:dL_ang}
\end{align}
Finally, using equation \eqref{eq:mB}, one can find the apparent magnitude that a supernova Ia would have at $d_L^{\left\lbrace \alpha,\theta \right\rbrace}(z_{\rm{bao}})$:
\begin{align}
m^{\left\lbrace \alpha,\theta \right\rbrace}_B (z_{\rm{bao}}) & =  5 \log_{10} \left[ \frac{d_L^{\left\lbrace \alpha,\theta \right\rbrace}(z_{\rm{bao}})}{1 \rm{Mpc}} \right] + 25 + M_B \,. \label{eq:mB_bao}
\end{align}
We can then directly compare $m^{\left\lbrace \alpha,\theta \right\rbrace}_B (z_{\rm{bao}}) $ with the $\overline m_{Ba}(z_{\rm{bao}})$ of equation~\eqref{eq:mB_logz}.

Equations \eqref{eq:mB_cosmography} and \eqref{eq:mB_bao} describe well our approach. No fiducial model is assumed and the cosmographic expansion is only used for the local SNe-1. This means that all information at distances greater than $\sim 600$ Mpc (redshift $z=0.15$) will go directly in $M_B$.

\section{Cosmic ladder analyses} \label{analyses}

Here, we will discuss three analyses that can be performed using the cosmic distance ladder:
\begin{enumerate}[label=\arabic*.,leftmargin=.5\parindent]
\item using only SNe Ia and BAO---$r_d h$ calibration, 
\item using SNe Ia, BAO and a CMB prior on $r_d$---the inverse distance ladder,
\item using SNe Ia, BAO and a local astro-prior on $M_B$---the extended distance ladder. 
\end{enumerate}

\subsection{SNe + BAO} \label{subsec:flat_case}

We have seen in equation \eqref{eq:mB_cosmography} that $M_B$ and $H_0$ are completely degenerate, that is, it is not possible to constrain them using only (supernova) apparent magnitude data. It is then useful to rewrite equation \eqref{eq:mB_cosmography} as follows:
\begin{align}
m^{\rm cg}_B & =  5 \log_{10} \left [\frac{c z H_0^{-1}}{\text{1 Mpc}} \, f  (z,q_0 ) \right] + 25 + M_B \nonumber \\
&= 5 \log_{10} \left [\frac{c z}{100 \ \rm{km/s}} \, f  (z,q_0 ) \right] + 25 + \hat{M}_B \,,
\end{align}
where we introduced the new variable $\hat{M}_B = M_B - 5 \log_{10}h$, which can be  constrained by supernova data.

We can then apply the same change of variable to equation \eqref{eq:mB_bao} so that:
\begin{align}
m^{\left\lbrace \alpha, \theta \right\rbrace}_B  = 5 \log_{10}\left[ \frac{d^{\left\lbrace \alpha, \theta \right\rbrace}_L}{r_d} \right] + 5 \log_{10} \frac{r_d h}{1 \rm{Mpc}}
 + 25 + \hat{M}_B \,,  \label{eq:mB2_rdh}
\end{align}
where the term $d^{\left\lbrace \alpha, \theta \right\rbrace}_L / r_d$ depends only on $\left\lbrace \alpha_\perp, \theta \right\rbrace$.
It is then clear that SNe-1 can constrain $\hat{M}_B$ and SNe-2 + BAO can constrain the combination $r_d h$ \citep{Heavens:2014rja,Verde:2016ccp}.

We will consider a 4-d posterior on the parameters $q_0$, $\hat{M}_B$, $\beta_i$, $r_d h$:
\begin{equation}
\mathcal{P}(q_0,\hat{M}_B,\beta_i,r_d h)  \propto \mathcal{L}_{\text{sne}}(q_0,\hat{M}_B,\beta_i,r_d h) \,\mathcal{L}_{\text{bao}}(\beta_i) \,,
 \label{eq:pos_rdh}
\end{equation}
where $\beta_i$ represents $\alpha^i_{\perp}$ or $\theta_i$ and we adopted improper flat priors.
The Gaussian distributions $\mathcal{L}(\theta) \propto e^{-\chi^2(\theta)/2} $ are:
\begin{gather}
\chi^2_{\text{bao}} = \left\lbrace \beta^{\text{obs}}_{i} - \beta_i \right\rbrace \Sigma^{-1}_{\beta ,ij} \left\lbrace \beta^{\text{obs}}_{j} - \beta_j \right\rbrace \,, \label{eq:chi2bao_rdh} \\
\chi^2_{\text{sne}} = \left\lbrace m_{Ba}^{\text{obs}} - F(z_a)\right\rbrace \Sigma^{-1}_{\text{sn},ab} \left\lbrace m_{Bb}^{\text{obs}} - F(z_b)\right\rbrace \,, \label{eq:chi2sne_rdh}
\end{gather}
where
\begin{equation}
F(z)=\left\lbrace
\begin{array}{ll}
m_B^{\rm cg}(z) \phantom{ciaociao}  0.023 \le z \le 0.15 \\
m_B^{\left\lbrace \alpha,\theta \right\rbrace}(z)  \phantom{ciacoo} 0.15 < z
\end{array}\,.
\right.
\end{equation}
Note that $\Sigma_{\text{sn}}$ is the supernovae covariance matrix after the binning process (which includes correlations between SNe-1 and SNe-2), see Section~\ref{sec:bin}. We have also analyzed the case in which correlations between SNe-1 and SNe-2 are neglected, see the disjoint distance ladder discussed in Appendix~\ref{ap:disjoint}.

Finally, we will also use SNe Ia and BAO data to constrain the degeneracy between $r_d$ and $H_0$, without any prior on $r_d$ or $H_0$. This will offer a useful way to compare our results with the CMB analysis from Planck 2018 \citep{Aghanim:2018eyx}.
To this end we will use:
\begin{equation}
\mathcal{P}(H_0,q_0,M_B,\beta_i,r_d) \!  \propto \! \mathcal{L}_{\text{sne}}(H_0,q_0,M_B,\beta_i,r_d) \mathcal{L}_{\text{bao}}(\beta_i) ,
 \label{eq:pos_rdh2}
\end{equation}
where the apparent magnitudes given in equations \eqref{eq:mB_cosmography} and \eqref{eq:mB_bao} are used.

\subsection{SNe + BAO + CMB prior}\label{sub:cmb_case}

We now discuss the so-called ``inverse distance ladder'' \citep[see, e.g.,][]{Macaulay:2018fxi,Feeney:2018mkj}.
BAO needs to be calibrated in order to provide a pure geometrical distance, that is, a standard ruler.
In the distance ladder jargon, we anchor the ladder at CMB via a prior on $r_d$.
Here, we adopt the prior of equation~\eqref{cmbprior}.
Thanks to this prior, the degeneracy between $H_0$ and $r_d$ is broken so that we can consider the parameters $H_0$, $q_0$, $M_B$, $\beta_i$, $r_d$ via the 5-d posterior:
\begin{align}
\nonumber \mathcal{P}(H_0,q_0,M_B,\beta_i,r_d)  \propto & \, \mathcal{L}_{\text{sne}}(H_0,q_0,M_B,\beta_i,r_d) \\
\times & \,  \mathcal{L}_{\text{bao}}(\beta_i)\, \mathcal{L}_{\text{cmb}}(r_d) \,,
 \label{eq:pos_H0}
\end{align}
where equations \eqref{eq:chi2bao_rdh} and \eqref{eq:chi2sne_rdh} are used, $\mathcal{L}_{\text{cmb}} \propto  e^{-\chi^2_{\rm{cmb}}/2}$ and
\begin{gather}
\chi^2_{\text{cmb}} = \frac{(r_d^{\text{obs}}-r_d)^2}{\sigma_d^2} \,. \label{eq:chi2cmb}
\end{gather}
For this analysis the apparent magnitudes given in equations~\eqref{eq:mB_cosmography} and~\eqref{eq:mB_bao} are used.

Finally, it is important to note that, although the CMB prior of equation~\eqref{cmbprior} corresponds to the analysis based on the $\Lambda$CDM model, this does not ensure that the $H_0$ value that we will obtain is in agreement with CMB determinations, see \citet{Macaulay:2018fxi}.

\subsection{SNe + BAO + astro-prior}\label{sub:local_case}

As discussed in Section~\ref{astrosec},
in \cite{Camarena:2019moy} we showed that it is possible to compress all the local information, provided by Cepheids and geometrical distances, in an informative prior on $M_B$, the  astro-prior of equation~\eqref{priorM}.

Here, we use the astro-prior in order to anchor the distance ladder at the other extremum, the local universe.
In this case the prior breaks the degeneracy between $H_0$ and $M_B$ and the ladder propagates this calibration via the BAO measurements into a model-independent determination of both $r_d$ and $H_0$. This is the extended distance ladder.

As in the previous Section, we will consider the parameters $H_0$, $q_0$, $M_B$, $\beta_i$, $r_d$ via the posterior: 
\begin{align}
\nonumber \mathcal{P}(H_0,q_0,M_B,\beta_i,r_d)  \propto & \, \mathcal{L}_{\text{sne}}(H_0,q_0,M_B,\beta_i,r_d) \\
\times & \,  \mathcal{L}_{\text{bao}}(\beta_i) \, \mathcal{L}_{\rm astro}(M_B) \,,
 \label{eq:pos_rd}
\end{align}
where the likelihoods are defined according to:
\begin{gather}
\chi^2_{\rm astro} = \frac{(M_B^{\text{loc}}-M_B)^2}{\sigma_{B}^2} \,, \label{eq:chi2H0}
\end{gather}
and equations \eqref{eq:chi2bao_rdh} and \eqref{eq:chi2sne_rdh}.

\section{Results}\label{sec:resul}

\begin{figure}
\centering
\includegraphics[width=\columnwidth]{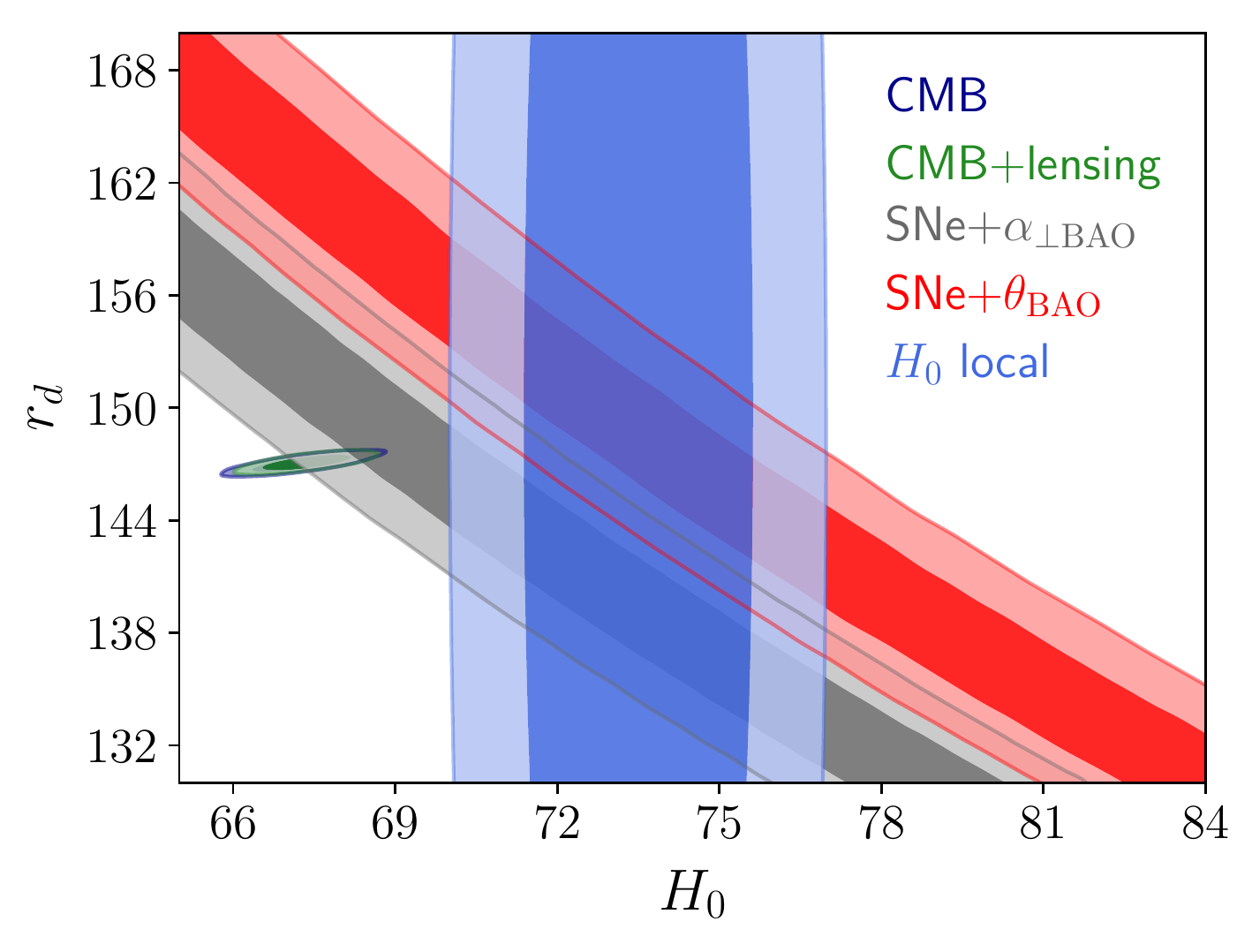}
\caption{$1$- and $2\sigma$ marginalized constraints on $r_d$ and $H_0$ using our cosmic distance ladder (only SNe and BAO). Gray for anisotropic BAO (perpendicular $\alpha_{\perp \rm BAO}$) and red for angular BAO  ($\theta_{\rm BAO}$). The small ellipses show the constraints by Planck 2018  \citep{Aghanim:2018eyx}: CMB and CMB+lensing represent results from TT+TE+EE+lowE and TT+TE+EE+lowE+lensing analyses, respectively. The blue band shows the local determination of $H_0$ by \citet{Reid:2019tiq}. }
\label{fig:flat_case}
\end{figure}

\begin{table*}
\begin{center}
\renewcommand{\arraystretch}{1.5}
\begin{tabular}{|l|>{\centering\arraybackslash}p{2.5cm}|>{\centering\arraybackslash}p{1.2cm}|>{\centering\arraybackslash}p{1.1cm}|>{\centering\arraybackslash}p{1.1cm}|>{\centering\arraybackslash}p{1.1cm}|>{\centering\arraybackslash}p{1.1cm}|}
\hline
\multirow{2}{*}{Type of analysis} & \multirow{2}{*}{\makecell{Constraint\\ $[$Mpc or km/s/Mpc$]$}} & \multicolumn{5}{c|}{Comparison with (Index of Inconsistency)}  \\ \cline{3-7} 
 &  & $r_d h$ from Planck  & $r_d$ from Planck   & $H_0$ from Planck  &  $H^{\rm Re19}_0$ (SH0ES) &  $H^{\rm CM}_0$ (SH0ES)   \\
 \hline
Section~\ref{subsec:flat_case}: SNe+$\alpha_{\perp \rm{BAO}}$ & $r_d h = 102.56 \pm 1.87$ & $1.7 \sigma$ & - & - & - & -  \\ \hline
Section~\ref{subsec:flat_case}: SNe+$\theta_{\rm{BAO}}$ & $r_d h = 109.37 \pm 2.09$ & $4.5 \sigma$ & - & - & -& -  \\ \hline 
Section~\ref{sub:cmb_case}: SNe+$\alpha_{\perp \rm{BAO}}+r_d^{\rm{cmb}}$ & $H_0 =  69.71 \pm 1.28$ & - & - & $1.7 \sigma$ & $2.0 \sigma$ & $2.7 \sigma$ \\ \hline
Section~\ref{sub:cmb_case}: SNe+$\theta_{\rm{BAO}}+r_d^{\rm{cmb}}$ & $H_0 = 74.36 \pm 1.42$ & - & - & $ 4.6 \sigma$ & $ 0.4 \sigma$ & $ 0.4 \sigma$ \\ \hline 
Section~\ref{sub:local_case}: SNe+$\alpha_{\perp \rm{BAO}}+M_B$ & \makecell{$r_d = 136.18 \pm 3.03$\\ $ H_0 = 75.32 \pm 1.68$} & - & $3.6 \sigma$ & $4.5 \sigma$ & $0.8 \sigma$ & $0.01 \sigma$\\ \hline
Section~\ref{sub:local_case}: SNe+$\theta_{\rm{BAO}}+M_B$ & \makecell{$r_d = 145.20 \pm 3.36$\\ $ H_0 = 75.36 \pm 1.68$} & - & $ 0.6 \sigma$ & $4.5 \sigma$ & $0.9 \sigma$ & $0.004 \sigma$\\ \hline
\end{tabular}
\caption{Constraints from our cosmic distance ladder according to the three analyses described in Section~\ref{analyses}.
We compare our results with the corresponding values from Planck 2018 (TT+TE+EE+lowE+lensing) and the local determinations of $H_0$ given in equations~(\ref{reid19}-\ref{cm19}).
The Planck values are $r_d = 147.09 \pm 0.26 \text{ Mpc}$, $r_d h= 99.08 \pm 0.92$ Mpc and $H_0 = 67.36 \pm 0.54 \text{ km/s/Mpc}$.}
\label{tab:full_pos}
\end{center}
\end{table*}

We have performed the Bayesian analyses described in the previous Section.
We have sampled the parameter space with \texttt{emcee} \citep{ForemanMackey:2012ig}, an open-source Markov chain Monte Carlo (MCMC) sampler, and analyzed the  chains with \texttt{getdist} \citep{Lewis:2019xzd}.
In order to compare directly with CMB constraints, we used the MCMC chains from the Planck 2018 analysis \citep[][TT+TE+EE+lowE+lensing]{Aghanim:2018eyx} available at \href{http://www.esa.int/Planck}{esa.int/Planck}.

\begin{figure}
\centering
\includegraphics[width=.85 \columnwidth]{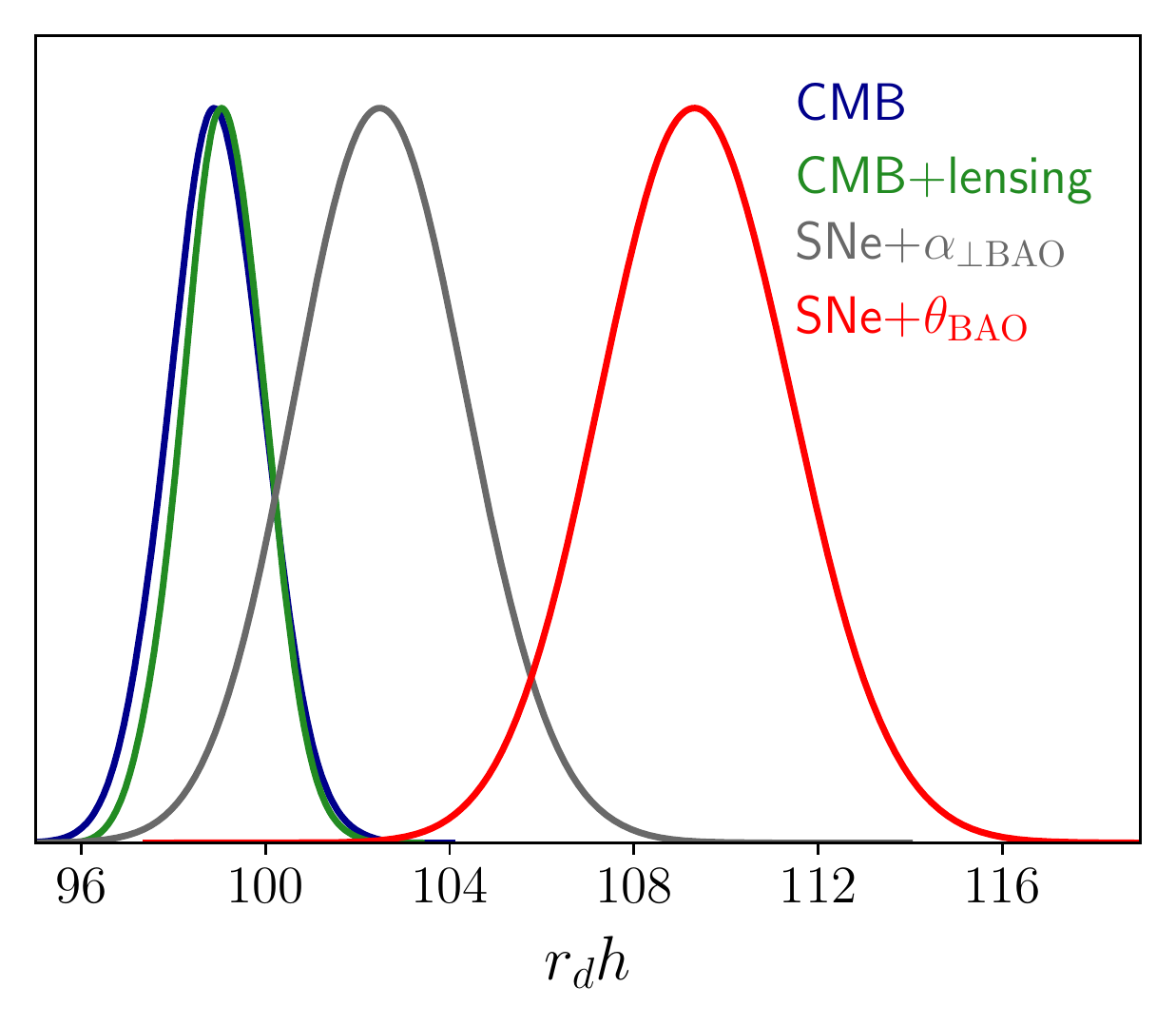}
\caption{Marginalized  posterior on $r_d h$ from our cosmic distance ladder (only SNe and BAO). Constraints using anisotropic BAO ($\alpha_{\perp \rm BAO}$, gray curve) seem to be in agreement with Planck 2018, while the ones from angular BAO ($\theta_{\rm BAO}$, red curve) show a $4.5 \sigma$ discordance with Planck 2018. See Table~\ref{tab:full_pos}.}
\label{fig:rdh_case}
\end{figure}

Our first result is shown in Figure~\ref{fig:flat_case} where we constrain the correlation direction between $r_d$ and $H_0$ using the posterior of equation \eqref{eq:pos_rdh2} (only SNe and BAO).
Even if it is not possible to constrain $r_d$ and $H_0$, this analysis is useful as it provides a qualitative insight of the data.
The negative correlation between $r_d$ and $H_0$ indicates that SNe+BAO data lead to lower (higher) values of $r_d$ when $H_0$ is shifted to higher (lower) values, 
as it was discussed in \citet{Evslin:2017qdn}.  
We can also see that the SNe$+\alpha_{\perp \rm{BAO}}$ determination is consistent with Planck 2018 results, which, however, is in tension with SNe$+\theta_{\rm{BAO}}$.
Also, although SNe+$\alpha_{\perp \rm{BAO}}$ shows agreement with Planck 2018, the inclusion of a prior that anchors the ladder could change this, as we will see below.
In order to quantify the tension it is useful to consider the posterior of equation \eqref{eq:pos_rdh}, where the degeneracy is removed by considering the combination $r_d h$, as originally proposed in \citet{Heavens:2014rja}. The result is given in Figure~\ref{fig:rdh_case}. Using now the Index of Inconsistency of \cite{Lin:2017bhs}, we find a $4.5 \sigma$ tension between Planck 2018 and angular BAO (SNe$+\theta_{\rm{BAO}}$), while Planck 2018 and anisotropic BAO are in good agreement ($1.7 \sigma$ tension).
See Table~\ref{tab:full_pos} for the numerical summary.

\begin{figure}
\centering
\includegraphics[width=.85 \columnwidth]{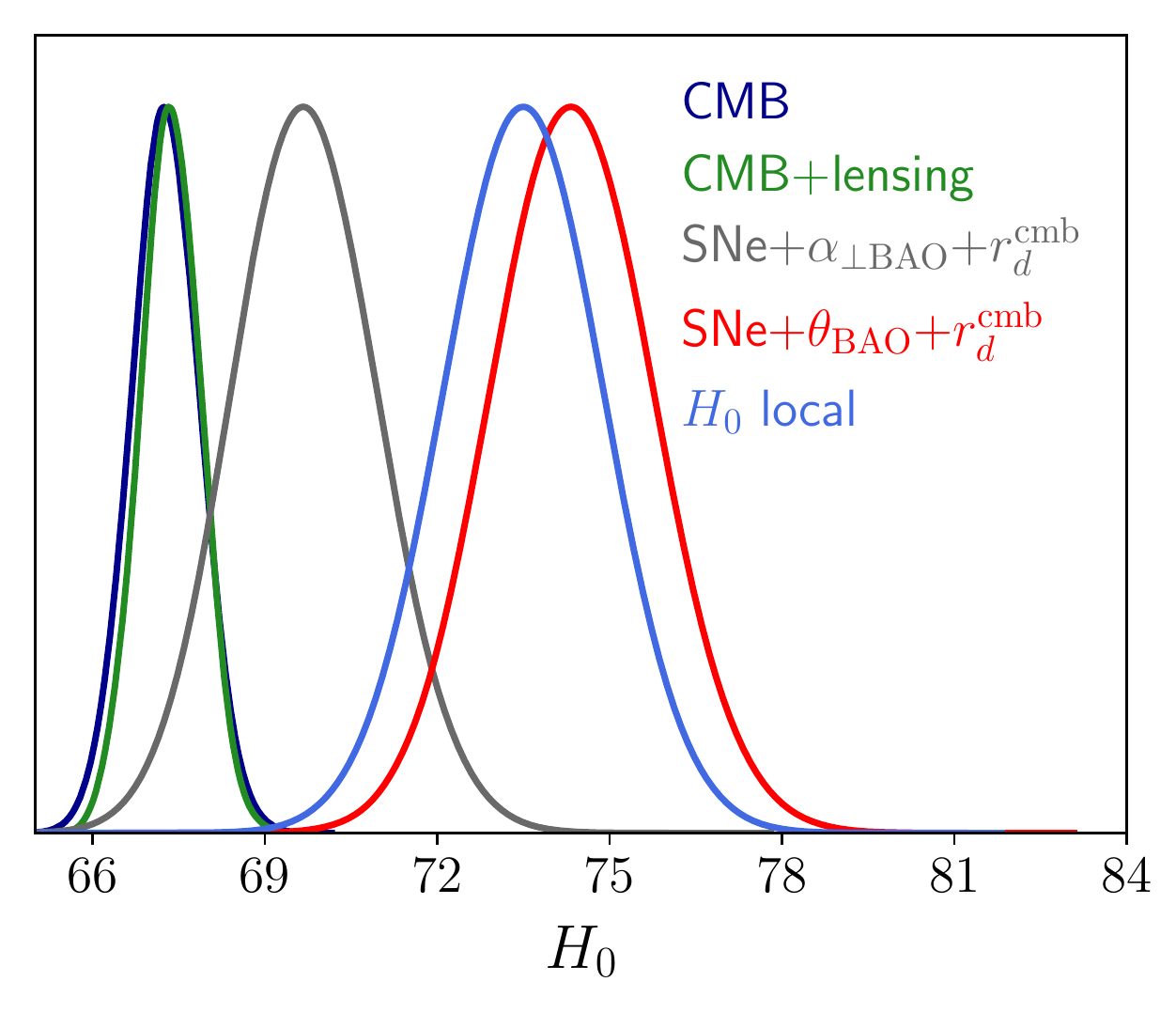}
\caption{Marginalized $H_0$ posterior obtained from the analysis using SNe, BAO and  CMB prior on $r_d$. Here, angular $\theta_{\rm BAO}$ (red curve) data supports the local determination and are at $4.6 \sigma$ tension with the value from Planck 2018. On the other hand, $\alpha_{\rm BAO}$ (gray curve) gives a value between the Planck 2018 result and the constraint from the local cosmic distance ladder. See Table~\ref{tab:full_pos}.}
\label{fig:H0_case}
\end{figure}

Next, we can anchor the cosmic ladder at the CMB epoch with a prior on $r_d$, that is, we will now consider the posterior of equation~\eqref{eq:pos_H0}.
The result is showed in Figure~\ref{fig:H0_case}. We do not report results on $r_d$ as they are prior dominated.
As expected from the analysis of Figure~\ref{fig:flat_case}, even if we use a CMB prior on $r_d$ from Planck, still the $\theta_{\rm{BAO}}$ data
gives a higher value of $H_0$, in perfect agreement with the local determination by \cite{Reid:2019tiq}, but at $4.6 \sigma$ tension with the Planck 2018 analysis.
On the other hand, using $\alpha_{\perp \rm{BAO}}$ we obtain a value of $H_0$ between the value from Planck 2018 and the local determination. See Table~\ref{tab:full_pos}.

\begin{figure}
\centering
\includegraphics[width= .85 \columnwidth]{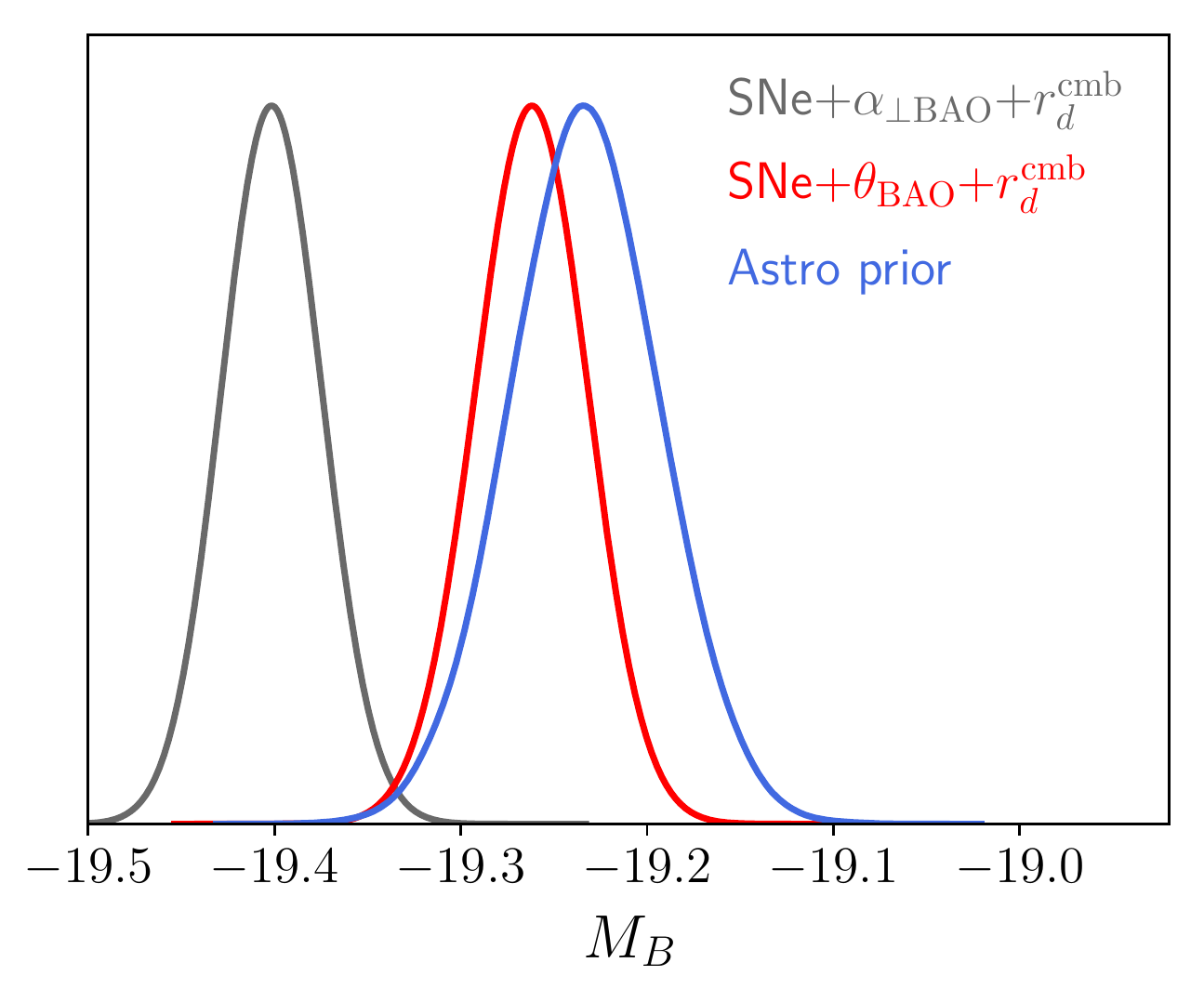}
\caption{Priors on the supernova absolute magnitude $M_B$.
The astro-prior comes from local physics while the CMB-priors from early-universe physics.}
\label{Mbees}
\end{figure}

From the posterior of equation~\eqref{eq:pos_H0} we can also obtain the so-called ``CMB-prior'' on $M_B$ by marginalizing over the other parameters:
\begin{align} \label{CMBprior}
M_B^{\alpha} = -19.401 \pm 0.027 
\qquad \quad
M_B^{\theta} = -19.262 \pm 0.030 \,. 
\end{align}
While the prior relative to angular BAO agrees with the astro-prior of equation \eqref{priorM} at $0.6 \sigma$, the one relative to anisotropic BAO disagrees at the $3.4 \sigma$ level.
Figure~\ref{Mbees} shows the astro-prior together with the CMB-priors.

\begin{figure}
\centering
\includegraphics[width= \columnwidth]{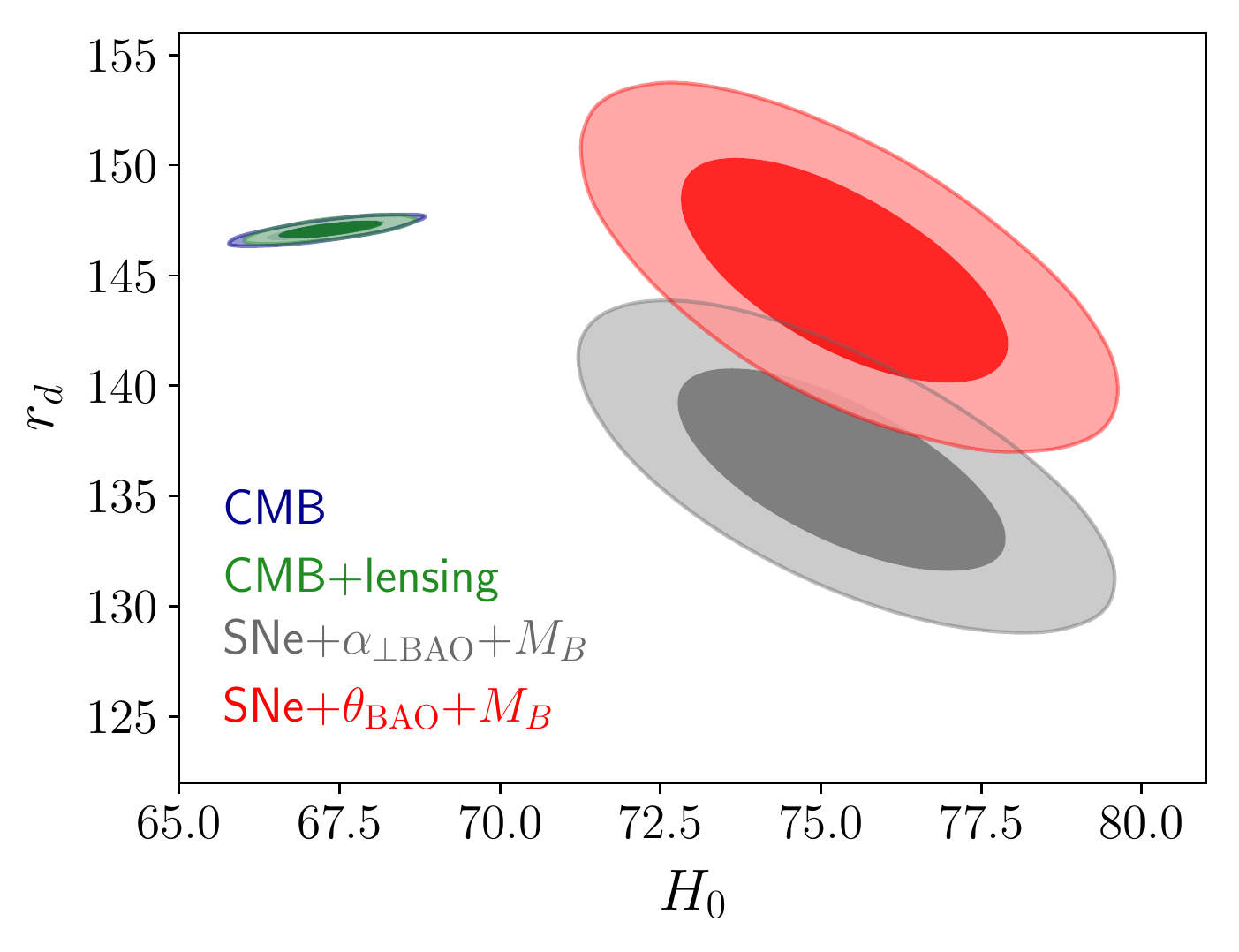}
\caption{$1$- and $2\sigma$ marginalized constraints on $r_d$ and $H_0$ when using SNe, BAO and the astro-prior on $M_B$. The use of the local astro-prior leads to a higher value of $H_0$ for both BAO cases. However, the value obtained for $r_d$ depends on the BAO data considered.
See Table~\ref{tab:full_pos}.}
\label{fig:MB_case}
\end{figure}

Finally, we can anchor the cosmic ladder at our present-day local universe, that is, we will now discuss the posterior of equation~\eqref{eq:pos_rd}, which uses the prior on the supernova absolute magnitude $M_B$ that was obtained in \cite{Camarena:2019moy}.
Results are showed in Figure~\ref{fig:MB_case} and Table~\ref{tab:full_pos}.
As expected from the analysis of Figure~\ref{fig:flat_case} and the discussion of Section~\ref{locdet}, anisotropic and angular BAO data give similar constraints on $H_0$, both in agreement with the value obtained in \citet{Camarena:2019moy} and at $4.5 \sigma$ tension with the value of the Hubble constant from Planck. 
On the other hand, the value obtained for the sound horizon at drag epoch $r_d$ depends on the BAO data used. Angular BAO data provides an $r_d$ in agreement with Planck 2018, but anisotropic measurements of BAO lead to an $r_d$ value that is at $3.6 \sigma$ tension with CMB.

The posteriors of the previous analyses all feature the deceleration parameter $q_0$.
We can then obtain and compare the marginalized posteriors on $q_0$. The result is given in 
Table~\ref{tab:q0_pos}.
It is clear that the analyses give almost the same $q_0$ constraint, in agreement with the value  obtained in \citet{Camarena:2019moy} (see also \citealt{Feeney:2017sgx}), where only SNe Ia and the astro-prior on $M_B$ were used. This shows that the local value of $q_0$ does not depend on how the cosmic ladder is anchored.

\begin{table}
\begin{center}
\renewcommand{\arraystretch}{1.5}
\begin{tabular}{|l|l|}
\hline
Type of analysis  &  Constraint on $q_0$  \\ \hline
SNe+$\alpha_{\perp \rm{BAO}}$ & $ q_0 = -1.08_{-0.29}^{+0.29}$ \\ \hline
SNe+$\theta_{\rm{BAO}}$ & $ q_0 = -1.11_{-0.29}^{+0.29}$  \\ \hline \hline
SNe+$\alpha_{\perp \rm{BAO}}+r_d^{\rm{cmb}}$ & $ q_0 = -1.09_{-0.29}^{+0.29}$   \\ \hline
SNe+$\theta_{\rm{BAO}}+r_d^{\rm{cmb}}$ & $ q_0 = -1.11_{-0.29}^{+0.29}$ \\ \hline \hline
SNe+$\alpha_{\perp \rm{BAO}}+M_B$ & $ q_0 = -1.08_{-0.29}^{+0.29}$ \\ \hline 
SNe+$\theta_{\rm{BAO}}+M_B$ & $ q_0 = -1.11_{-0.29}^{+0.29}$ \\ \hline
\end{tabular}
\caption{Determination of the deceleration parameter $q_0$ for all the cases considered in this work. Note that for all cases the standard $\Lambda$CDM value $q_0 \approx -0.55$ disagrees with the observational values at the $2\sigma$ level.}
\label{tab:q0_pos}
\end{center}
\end{table}

\section{Comparison with previous inverse distance ladders}\label{sec:discu}

\begin{figure*}
\centering
\includegraphics[width=.85\textwidth]{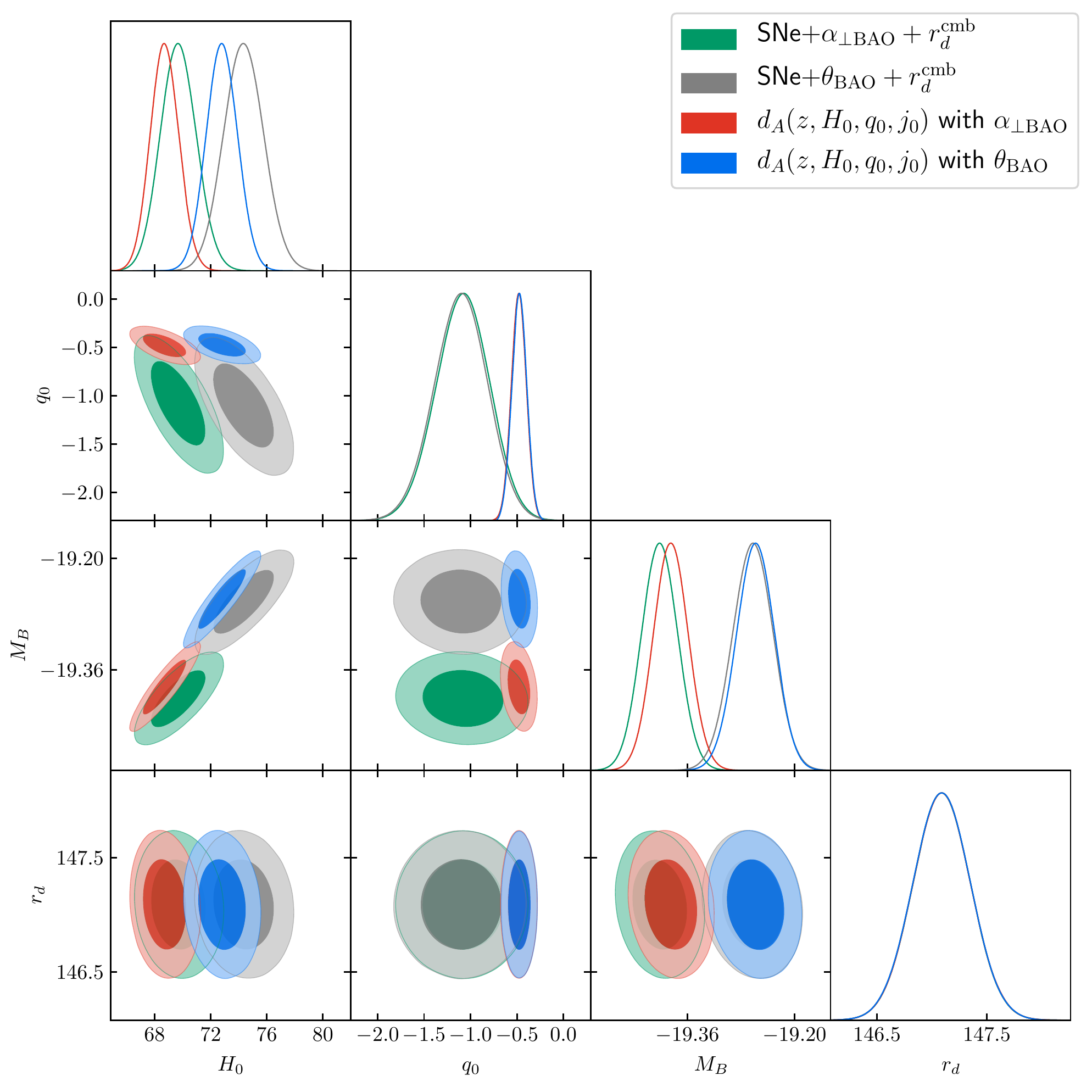}
\caption{$1$- and $2\sigma$ marginalized constraints on $H_0$, $q_0$, $M_B$ and $r_d$ when the cosmic ladder is built using the distance-duality method (SNe+$\alpha_{\perp \rm{BAO}}+r_d^{\rm{cmb}}$ and SNe+$\theta_{\rm{BAO}}+r_d^{\rm{cmb}}$) and the cosmographic method ($d_A(z,H_0,q_0,j_0)$ with $\alpha_{\perp \rm{BAO}}$ and $\theta_{\rm{BAO}}$). See Section~\ref{sec:discu}.}
\label{fig:our-inv_vs_pure-cosmo}
\end{figure*}

We will now compare our inverse distance ladder of Section~\ref{sub:cmb_case} with previous results.
The standard way to extend the cosmic ladder to high-redshift observations is via a cosmographic approach \citep[see, e.g.,][]{Macaulay:2018fxi,Feeney:2018mkj}; we call ``cosmographic method'' this way of building the comic ladder.
As discussed earlier, our  method uses the cosmographic expansion only at low redshifts ($z\le 0.15$) and exploits the distance-duality relation for calibrating BAO and SNe data; we name it the ``distance-duality method.''
In Figure~\ref{fig:our-inv_vs_pure-cosmo} we compare the comic ladders built with the cosmographic and distance-duality method. For the former, we adopted the method of \citet{Feeney:2018mkj}, considering a cosmographic expansion at second-order that covers  all distances, till $z \approx 2$. 
We have labelled the cosmographic method analyses as $d_A(z,H_0,q_0,j_0) + \alpha_{\perp \rm{BAO}}$ and $d_A(z,H_0,q_0,j_0) + \theta_{\rm{BAO}}$ for the anisotropic and angular BAO case, respectively.
From Figure~\ref{fig:our-inv_vs_pure-cosmo} one can note that the cosmographic method gives tighter constraints. This is expected as a single model is fitted to supernovae both at low and high redshift.
In particular, the degeneracy between $H_0$ and $q_0$ is broken.
However, for the very same reason, the cosmographic method forces correlation between $M_B$, and so $H_0$, and the shape of the luminosity-distance relation at $0\lesssim z \lesssim 2$.
In other words, $M_B$ and $H_0$ depend on physics beyond the low-redshift local universe, in particular on the properties of dark energy.
This explains why the posteriors on $q_0$ and $H_0$ differ. In particular, when using the cosmographic method, the posterior on $q_0$ is centered around the standard model value of $\approx -0.55$, in agreement with the findings of \citet{Macaulay:2018fxi}.
This clearly shows how high-$z$ supernovae affect the local parameters.
Our distance-duality method does not have this problem as low-redshift supernovae are used to obtain $H_0$ and $q_0$ and high-redshift supernovae to calibrate $M_B$ with BAO observations via the distance-duality relation.

\begin{figure*}
\centering
\includegraphics[width=0.85\textwidth]{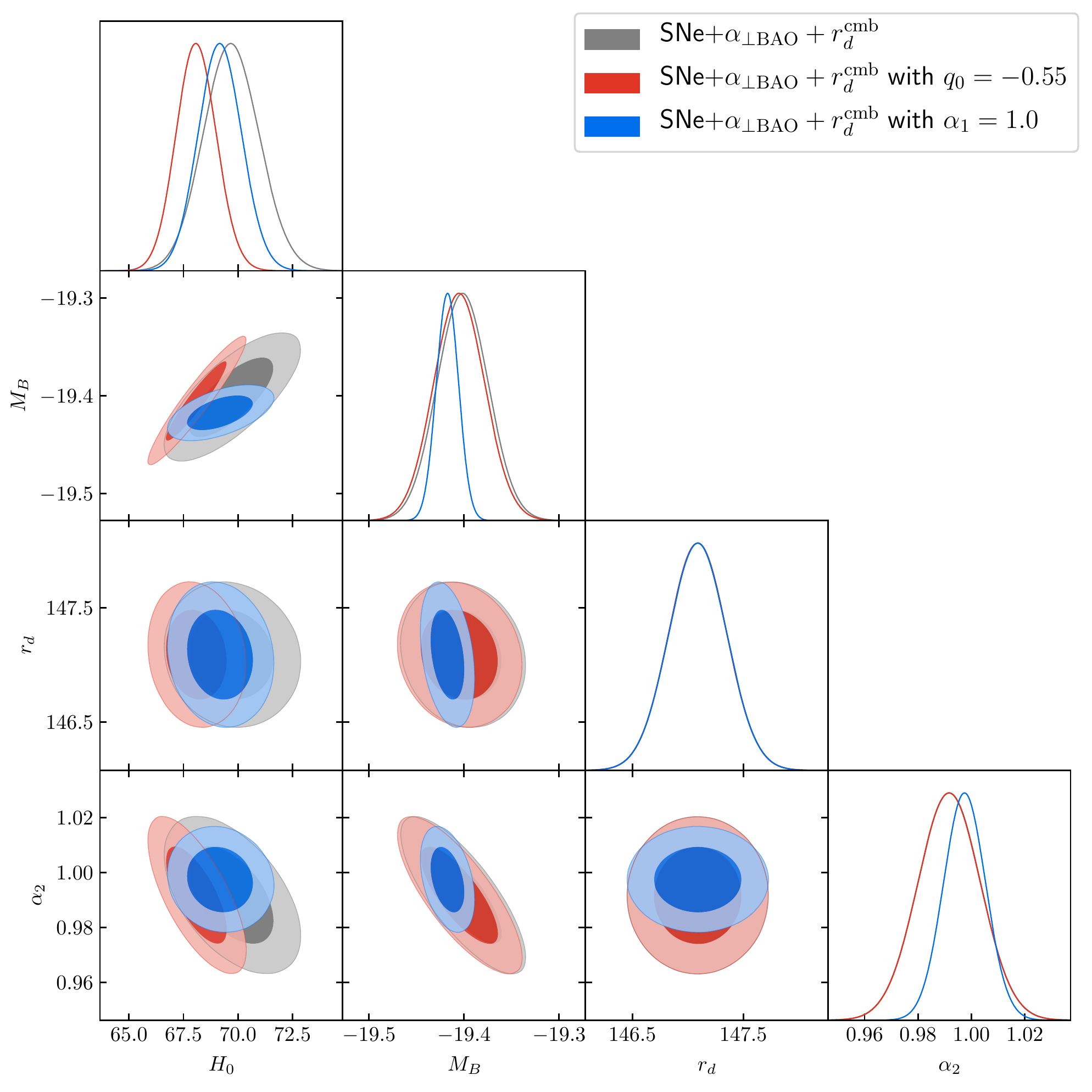}
\caption{$1$- and $2\sigma$ marginalized constraints on $H_0$, $M_B$, $r_d$ and $\alpha_2$ for three SNe+$\alpha_{\perp \rm{BAO}}+r_d^{\rm{cmb}}$ analyses: leaving $q_0$ and $\alpha_1$ free (as in Figure~\ref{fig:our-inv_vs_pure-cosmo}, green), fixing $q_0=-0.55$ and fixing $\alpha_1 = 1$.  See Section~\ref{sec:discu}.}
\label{fig:q0compare_ani}
\end{figure*}

It is important to note that other parameterizations besides the cosmographic expansion can be used in order to constrain $H_0$ \citep{Lemos:2018smw,Lyu:2020lwm,Bernal:2016gxb,Tutusaus:2018ulu}.
These approaches can be more flexible than the cosmographic one and are also model-independent.
However, some (often minimal) assumptions about the functional shape of $H(z)$ are needed.
For instance, \citet{Lemos:2018smw} uses parameterizations that can be understood as phenomenological extensions of $\Lambda$CDM, including, for example, in $H^2(z)/H^2_0$ terms such as $(1+z)^\epsilon$ or $\ln(1+z)$ which are expected to mimic a departure from~$\Lambda$.
Something similar happens in \citet{Tutusaus:2018ulu}, where data fits several `knots' that are interpolated by a piece-wise natural cubic spline that is then used to reconstruct $H(z)$.
Although these approaches  can be meaningfully used to reconstruct $H(z)$ (or even distances), 
our method differs from them since 
the calibration of supernovae with BAO observations is performed without having to reconstruct the luminosity distance.
Consequently,  no assumption is made regarding the functional form of the luminosity distances or the dynamical behavior of the universe beyond $z=0.15$.

In order to further clarify the advantages of our method, we have performed two more analyses of the case SNe+$\alpha_{\perp \rm{BAO}}+r_d^{\rm{cmb}}$:
one fixing the deceleration parameter to $q_0 =-0.55$ and another one with one of the BAO parameters fixed to unity ($\alpha_1 = 1$). See Figure~\ref{fig:q0compare_ani}.
Fixing $q_0$ affects the constraint on $H_0$, but not the calibration of $M_B$, which is calibrated via the distance-duality relation.
On the other hand, fixing $\alpha_1=1$ impacts the calibration of $M_B$ and to a smaller degree the determination of $H_0$. This means that the probes at $z > 0.15$ directly calibrate the absolute magnitude $M_B$. High redshift data does not bias our local determinations. This it is also discussed in Appendix~\ref{ap:disjoint}.

\section{Conclusions} \label{sec:conclu}

We have presented a new method to build the cosmic distance ladder, going from local astrophysical measurements to the CMB.
Instead of relying on a cosmography in order to model the luminosity distance and calibrate supernovae with BAO, we exploited the distance-duality relation $d_L = (1+z)^2 d_A$---valid if photon number is conserved and gravity is described by a metric theory.
The use of a cosmographic model forces correlation between $H_0$ and the shape of the luminosity-distance relation at $0\lesssim z \lesssim 2$. Therefore, it does not provide a local determination of $H_0$.
Our approach overcomes these issues as no model is used to calibrate supernovae with BAO. Thanks to this, one can directly identify the role of BAO measurements on the Hubble tension.

First, using the latest supernova and BAO observations, we have found that the combination $r_d h$ obtained from angular BAO measurements is in tension with the Planck determination at the $4.5 \sigma$ level. This strong result is rather robust as no CMB or local data is used.
On the other hand, using the anisotropic BAO measurements we found that $r_d h$ deviates from the Planck result by  $1.7 \sigma$. This inconsistency between angular and anisotropic BAO measurements should be investigated.

The latter inconsistency reflects also on the analysis that include a CMB prior on $r_d$.
The angular BAO constraint gives a posterior on $H_0$ in perfect agreement with the local determination by  \citet{Reid:2019tiq} but in strong tension with Planck ($4.6\sigma$).
When one uses the anisotropic BAO result one obtains a posterior that lies in the middle with respect to the local and CMB determination.
This is seen in a clearer way if one looks at the different calibrations of the supernova absolute magnitude.
When using angular BAO, the calibration of $M_B$ is consistent with the local one, but when using the anisotropic BAO there is tension.

The third analysis adopts the local astrophysical prior on $M_B$ so that the determinations of $H_0$ are, as expected, in strong tension with the CMB constraint.
However, this time, the constraint on $r_d$ from the angular BAO is in agreement with CMB and the one from anisotropic BAO in tension at the $3.6\sigma$ level.

Finally, we have found that, for all the cases, local supernovae ($z<0.15$) constrain the deceleration parameter to $q_0 \approx -1.10 \pm 0.29$, which is in tension with the standard model value ($\Omega_m = 0.3$, $\Omega_\Lambda = 0.7$) of $q_0 = -0.55$ at the $2\sigma$ level.
The same result was obtained from the analysis of low-redshift supernovae \citep{Camarena:2019moy}.
Note that in the $\Lambda$CDM model $ q_0 < -1 $ is equivalent to $\Omega_m < 0$, which could be interpreted as a local underdensity at $z <0.15$  \citep{Colgain:2019pck}.

Summarizing, we found a consistently low value of $q_0$ and strong inconsistency between angular and anisotropic BAO measurements, which are, or not, in agreement with CMB depending on the kind of analysis.
In order to solve this puzzle, a first step should be clarifying the tension between angular and anisotropic BAO measurements as this will help  understanding if new physics is required at the pre-recombination epoch and/or during the dark energy era.

\section*{Acknowledgements}
DC thanks CAPES for financial support.
VM thanks CNPq and FAPES for partial financial support.
This work also made use of the Virgo Cluster at Cosmo-ufes/UFES, which is funded by FAPES and administrated by Renan Alves de Oliveira.
The authors are grateful to Antonio Batista Brasil for his invaluable dedication and contributions to the PPGCosmo and PPGFis graduate programs and to the Center for Astrophysics and Cosmology Cosmo-ufes.


\bibliographystyle{mnras}
\bibliography{biblio}

\appendix


\section{Cosmographic expansion at the order $\mathcal{O}(\lowercase{z}^3)$}\label{ap:cosmography}

The cosmographic expansion is a Taylor expansion on redshift $z$ that allows one to write cosmological distances in a model-independent way.
Note, however, that this does not mean that its parameters do not contain cosmological information but rather that the parametrization of the distances is chosen in a model independent way.

We have used the cosmographic expansion in Section~\ref{locdet} in order to obtain the luminosity distance of SNe Ia in the range $0.023 < z < 0.15$.
We adopted a cosmographic expansion at the order $\mathcal{O}(z^2)$ because in this redshift range the weighted error from neglecting the higher-order correction is only 0.2\% \cite[see][figure~1]{Camarena:2019moy}.

Nevertheless, in order to assess the impact of this approximation, we performed the analysis of SNe+$\alpha_{\perp \rm{BAO}}+r_d^{\rm{cmb}}$ using the cosmographic expansion at the order $\mathcal{O}(z^3)$:
\begin{align}
f(z,q_0,j_0) \!=\!1 \!+\! \frac{1\!-\! q_0}{2}z \!-\! \frac{1-q_0-3q_0^2+j_0}{6}z^2 \!+\! O(z^3)  , \label{eq:mB_cosmography_2}
\end{align}
where the deceleration parameter $q_0$ and the jerk parameter $j_0$ are defined according to:
\begin{align}
q_0=-\left.\frac{\ddot a(t)}{H^2(t) a(t)}\right|_{t_0}
\qquad \quad
j_0=\left.\frac{\dddot a(t)}{H^3(t) a(t)}\right|_{t_0} \,.
\end{align}

\begin{figure}
\centering
\includegraphics[width= \columnwidth]{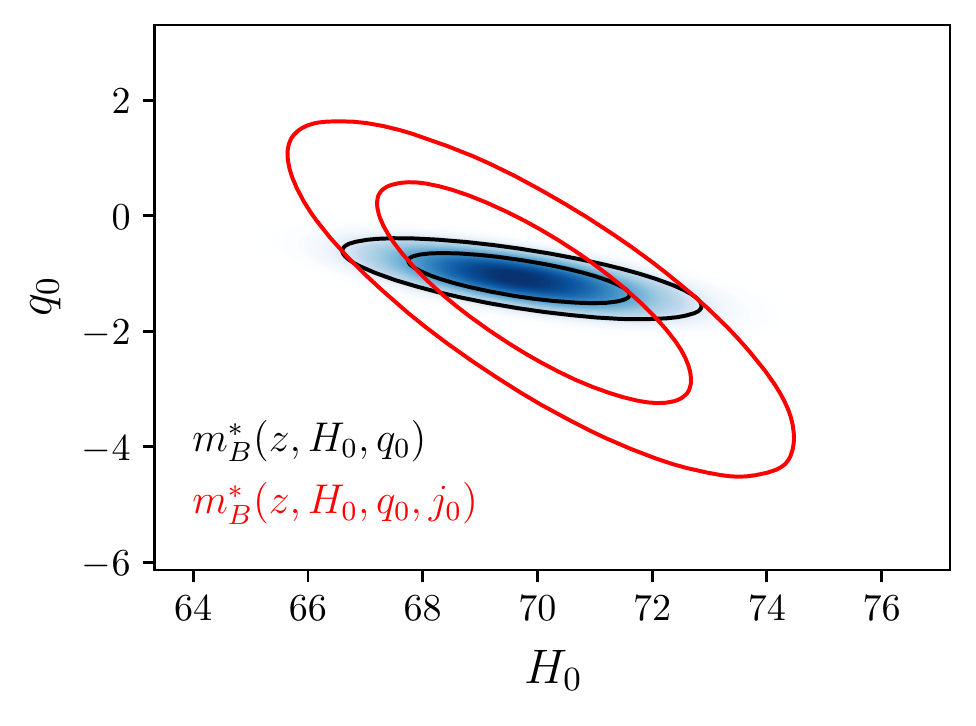}
\caption{$1$- and $2\sigma$ marginalized constraints on $H_0$ and $q_0$ for the case SNe+$\alpha_{\perp \rm{BAO}}+r_d^{\rm{cmb}}$ when using equation~\eqref{eq:mB_cosmography} (labelled as $m_B^*(H_0,q_0)$) and equation~\eqref{eq:mB_cosmography_2} (labelled as $m_B^*(H_0,q_0,j_0)$).
}
\label{fig:ap2}
\end{figure}

The comparison with the $\mathcal{O}(z^2)$ analysis of equation~\eqref{eq:mB_cosmography} is given in Figure~\ref{fig:ap2}.
The number of supernovae in the redshift range $0.023< z < 0.15$ is limited and the jerk parameter is poorly constrained, $j_0 = 11.80_{-37.17}^{+18.70}$. Consequently, also the constraints on $H_0$ and $q_0$ are weakened. The uncertainty on $H_0$ goes from 1.8\% to 2.5\%. 
Also, the bias on $H_0$ and $q_0$ due to neglecting the third order is $0.2 \sigma$ and $1 \sigma$, respectively.
This means that neglecting the third order does not affect the local constraint on $H_0$. However, as expected from correlations and the fact that $j_0$ is poorly constrained, the inclusion of the jerk parameter has an impact on the deceleration parameter.

\section{Our binning method vs Pantheon's binning}\label{ap:vsPantheon}

\begin{figure*}
\centering
\includegraphics[width=0.65\textwidth]{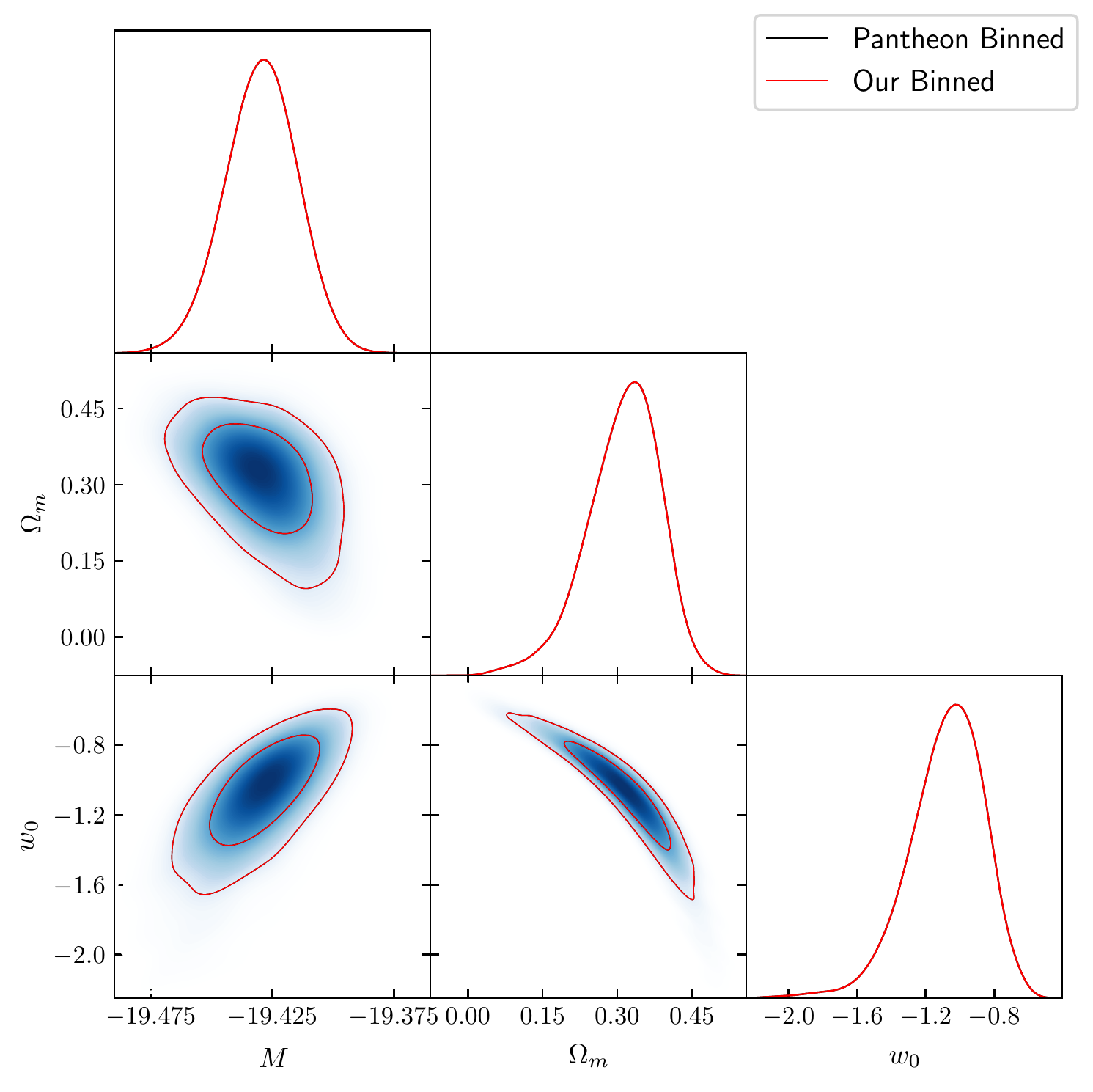}
\caption{$1$- and $2\sigma$ marginalized constraints on $w_0$, $\Omega_m$ and $M_B$ for $w$CDM model using our binned catalog and Pantheon binned catalog. The two catalogs produce the same results.}
\label{fig:ourbin_pantheon}
\end{figure*}

In order to validate our binning method of Sections~\ref{sec:bin}, we have produced, starting from the full Pantheon dataset, a binned catalog similar to one provided by \citet[][Appendix A]{Scolnic:2017caz}.
Here, we perform a Bayesian analysis of a $w$CDM model using these two compressed datasets.
We have used Montepython \citep{Audren:2012wb} and CLASS \citep{Blas:2011rf}, see Figure~\ref{fig:ourbin_pantheon}. 
The result shows that the two binned catalogs are indistinguishable and provide the same constraints over the cosmological parameters.
Note that here $M_B$ is not calibrated and degenerated with $\log_{10}H_0$.
We can conclude that the binning process of Section~\ref{sec:bin} is robust and that we do not expect  any bias.

\section{Disjoint ladder}\label{ap:disjoint}

\begin{figure*}
\centering
\includegraphics[clip,trim = 0 12.3cm 0 0,width=0.75\textwidth]{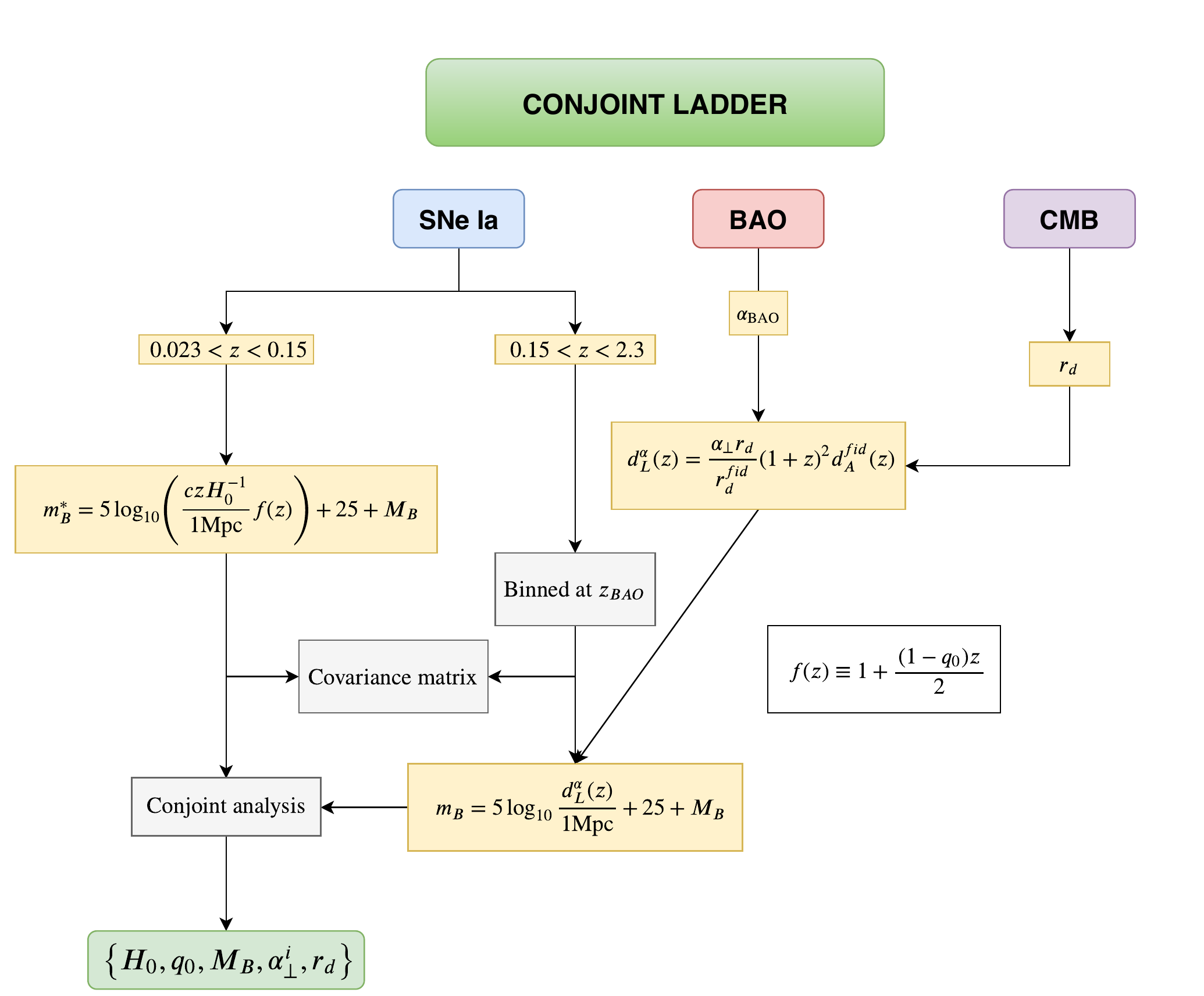}
\\
\includegraphics[clip,trim = 0 12.3cm 0 0,width=0.75\textwidth]{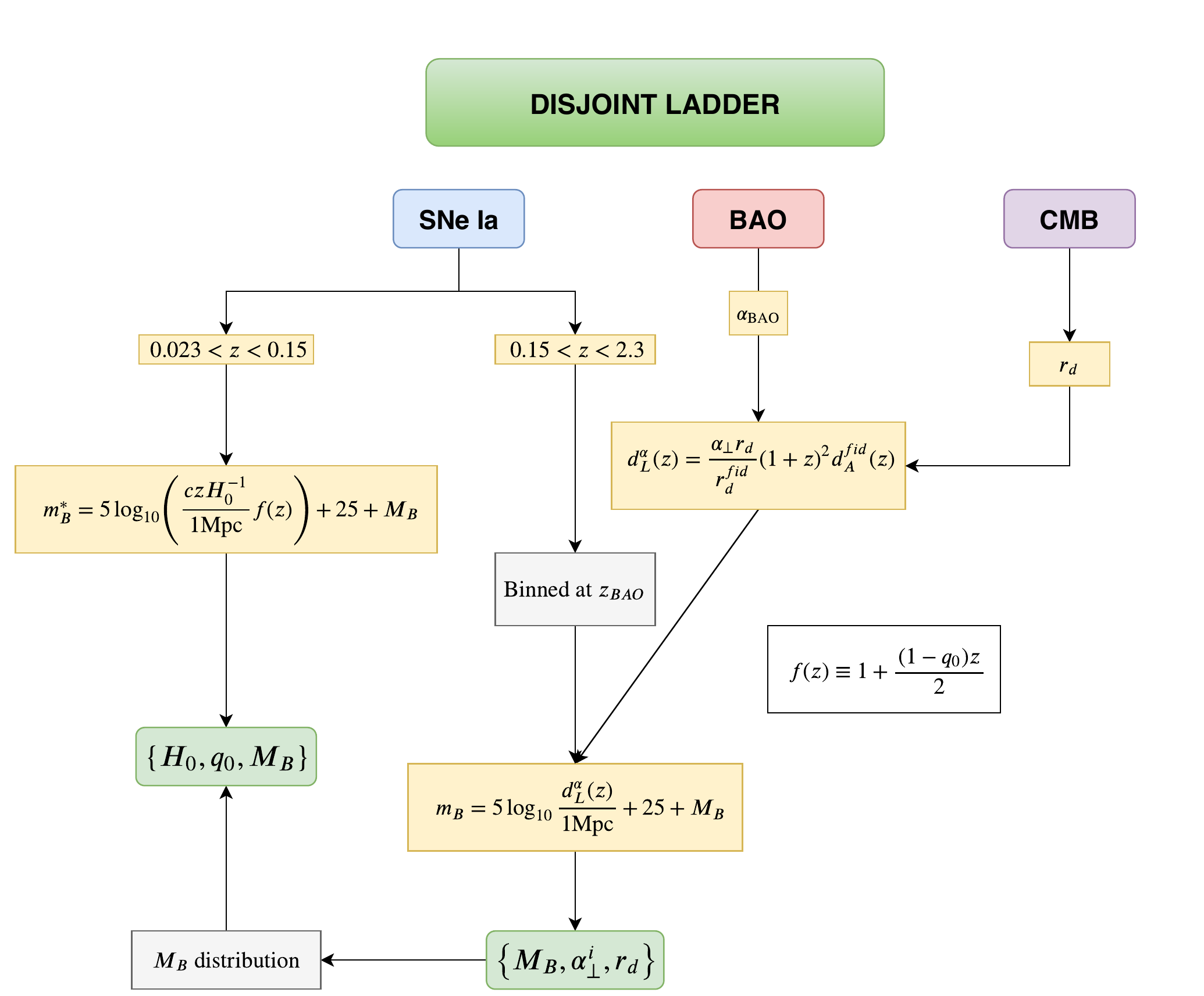}
\caption{Upper: Flowchart of the conjoint ladder for the case SNe+$\alpha_{\perp \rm{BAO}}+r_d^{\rm{cmb}}$ of Section~\ref{sub:cmb_case}. The conjoint analysis keeps correlations between SNe-1 and SNe-2, with the posterior defined by equation~\eqref{eq:pos_H0}.  Lower: Flowchart of the disjoint ladder as described in Appendix~\ref{ap:disjoint}. In this approach we neglect correlations among SNe-1 and SNe-2 so that only the calibration of $M_B$ connects SNe-2 to SNe-1.}
\label{fig:conjoint_disjoint}
\end{figure*}

SNe-1 and SNe-2 interact through the absolute magnitude $M_B$ and this cannot bias, for example, the local determination of $H_0$ via SNe-1. However, $H_0$ could be affected by the correlations between SNe-1 and SNe-2 present in the supernova covariance matrix.
In order to avoid this possible issue, we have also investigated the ``disjoint'' distance ladder, where correlations between SNe-1 and SNe-2 are neglected.
Here, we refer to the cosmic ladder discussed in the main text as to the ``conjoint'' distance ladder.

The disjoint distance ladder is built as described in Figure~\ref{fig:conjoint_disjoint}.
First, we calibrate $M_B$ using SNe-2, BAO data and the CMB prior:
\begin{align}
\nonumber \mathcal{P}_2(M_B,\alpha_{\perp}^i,r_d)  \propto  \mathcal{L}_{\text{sne-2}}(M_B,r_d,\alpha_{\perp}^i)  \mathcal{L}_{\text{cmb}}(r_d) \mathcal{L}_{\text{bao}}(\alpha_{\perp}^i) \,,
 \label{eq:fullpos_d1}
\end{align}
where the likelihoods for BAO and CMB are defined as in equations~\eqref{eq:chi2bao_rdh} and~\eqref{eq:chi2cmb}. The $\chi^2$ function for SNe-2 is defined according to:
\begin{equation}
\chi^2_{\text{sne-2}} = \left\lbrace m_{Bi}^{\text{obs}} - m_B(z_i)\right\rbrace \Sigma^{-1}_{\text{sn2},ij} \left\lbrace m_{Bj}^{\text{obs}} - m_B(z_j)\right\rbrace \,. \label{eq:chi2sne1}
\end{equation}
Then, we use the marginalized posterior on $M_B$ as a prior for SNe-1:
\begin{align}
\nonumber \mathcal{P}_1(H_0,q_0,M_B)  \propto \, & \mathcal{L}_{\text{sne-1}}(H_0,q_0,j_0,M_B) \\
\times & \int \mathcal{P}_{\text{2}}(M_B,\alpha_{\perp}^i,r_d) dr_d d\alpha_{\perp}^i \,,
 \label{eq:fullpos_d2}
\end{align}
where the corresponding $\chi^2$ function is:
\begin{equation}
\chi^2_{\text{sne-1}} = \left\lbrace m_{Bi}^{\text{obs}} - m^*_B(z_i)\right\rbrace \Sigma^{-1}_{\text{sn1},ij} \left\lbrace m_{Bj}^{\text{obs}} - m^*_B(z_j)\right\rbrace \,. \label{eq:chi2sne2}
\end{equation}
Note that equation~\eqref{eq:fullpos_d2} ensures that observables at $z> 0.15$ are only used to calibrate the supernova absolute magnitude. 

\begin{figure*}
\centering
\includegraphics[width=0.72\textwidth]{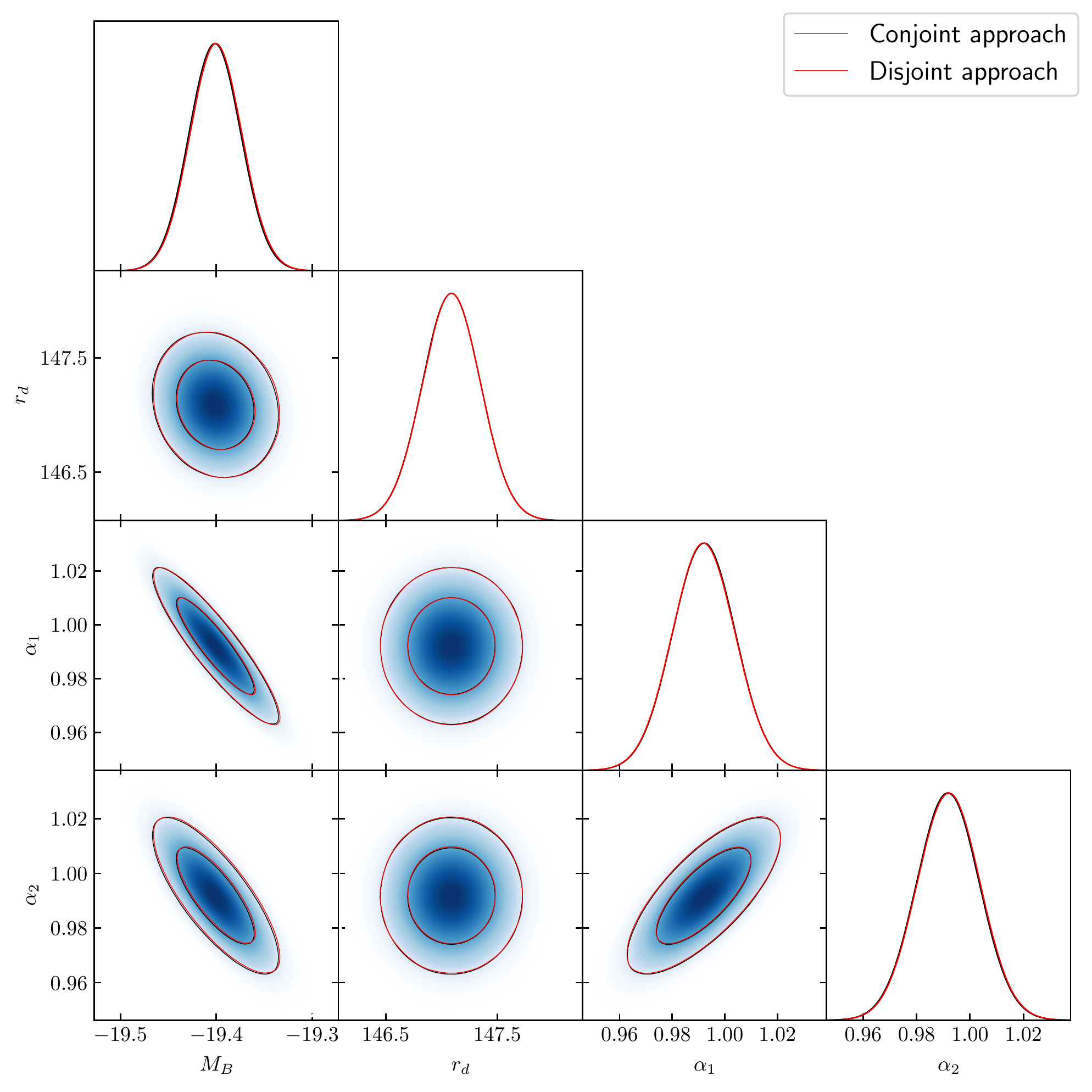}
\\
\includegraphics[width=0.58\textwidth]{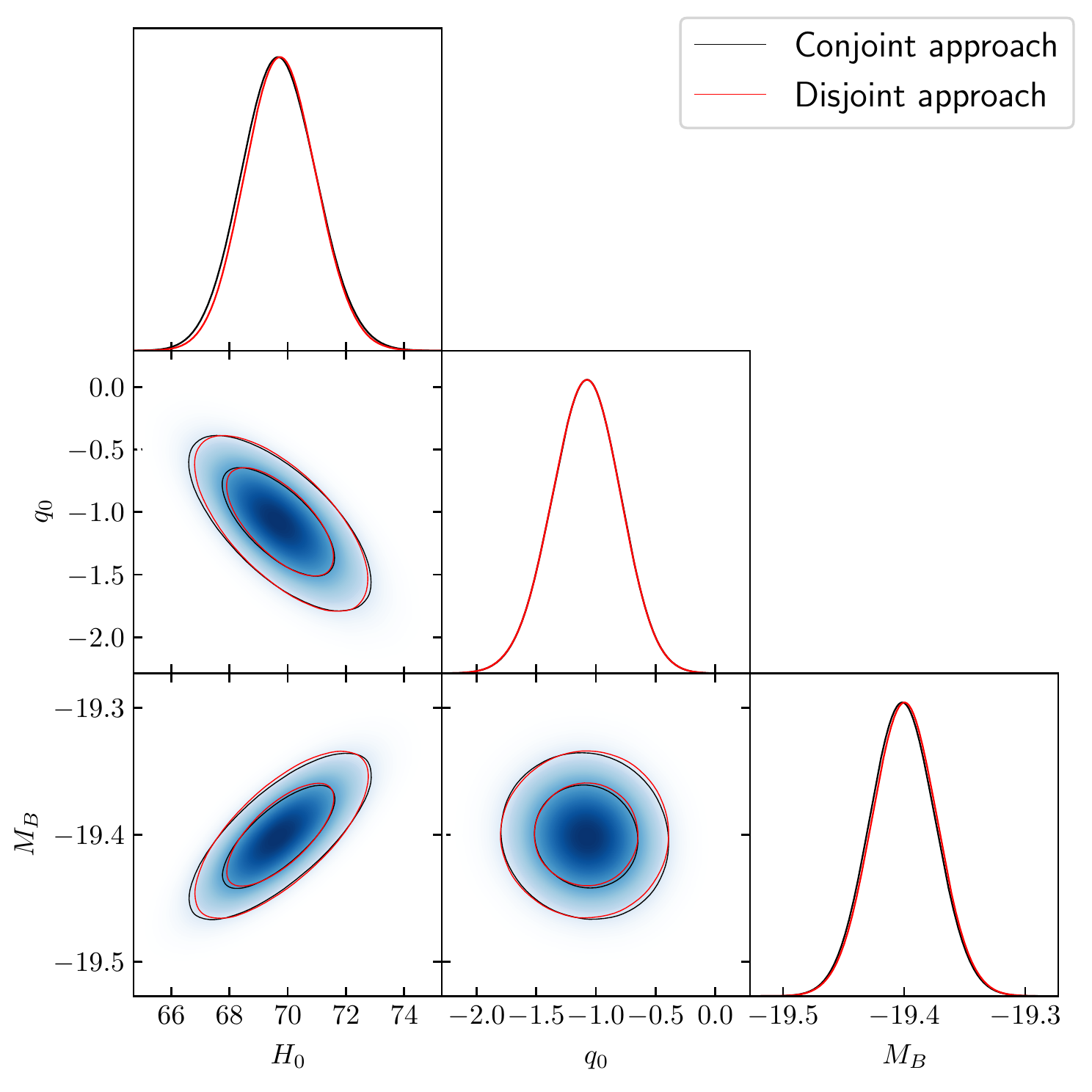}
\caption{Upper: $1$- and $2\sigma$ marginalized constraints on the parameters relative to SNe-2 for the disjoint and conjoint ladders. Lower: $1$-~and $2\sigma$ marginalized constraints on the parameters relative to SNe-1.}
\label{fig:conjoint_disjoint_distri}
\end{figure*}

In Figure \ref{fig:conjoint_disjoint_distri} we show the results of the analyses using the disjoint and conjoint ladders for the case SNe+$\alpha_{\perp \rm{BAO}}+r_d^{\rm{cmb}}$. The two analyses are almost indistinguishable.
We conclude that for present-day data correlations do not impact the analysis.

\bsp	
\label{lastpage}
\end{document}